\documentclass[useAMS,usenatbib,usegraphicx]{mn2e}

\def \arcsec{^{\prime\prime}}
\def \arcmin{^{\prime}}
\def\msun{{\rm M}_{\odot}}
\def\rsun{{\rm R}_{\odot}}

\def\rearth{{\rm R}_{\oplus}}
\def\teff{{\rm T}_{\rm eff}}
\def\sred{S_{\rm red}}

\usepackage{natbib,epsfig,graphicx,rotating,lscape,rotating,amssymb,color,subfigure}

\title[WTS: Masses and Radii of M-dwarf EBs]{Discovery and
  characterisation of detached M-dwarf eclipsing binaries in the WFCAM
  Transit Survey} \author[J. L. Birkby et al.]{Jayne
  Birkby$^{1,2}$\thanks{E-mail: jlb at ast.cam.ac.uk (JLB)}, Bas
  Nefs$^{2}$, Simon Hodgkin$^{1}$, G\'{a}bor Kov\'{a}cs$^{1}$, Brigitta
  Sip\H{o}cz$^{3}$, \newauthor David Pinfield$^{3}$, Ignas Snellen$^{2}$,
  Dimitris Mislis$^{1}$, Felipe Murgas$^{4}$, Nicolas
  Lodieu$^{4,5}$, \newauthor Ernst de Mooij$^{6}$, Niall
  Goulding$^{3}$, Patricia Cruz$^{7}$, Hristo Stoev$^{7}$, Michele
  Cappetta$^{8}$,\newauthor Enric Palle$^{4,5}$, David Barrado$^{7}$,
  Roberto Saglia$^{8}$, Eduardo Martin$^{9}$,
  Yakiv Pavlenko$^{10}$\\\\
  $^{1}$Institute of Astronomy, University of Cambridge, Madingley Road, Cambridge, CB3 0HA, UK\\
  $^{2}$Leiden Observatory, Leiden University, Niels Bohrweg 2, 2333 CA Leiden, The Netherlands\\
  $^{3}$Centre for Astrophysics Research, Science and Technology Research Institute, University of Hertfordshire, AL10 9AB, Hatfield, UK\\
  $^{4}$Instituto de Astrof\'isica de Canarias (IAC), C/ V\'ia  L\'actea s/n, E-38200 La Laguna, Tenerife, Spain\\
  $^{5}$Departamento de Astrof\'isica, Universidad de La Laguna (ULL), E-38205 La Laguna, Tenerife, Spain\\
  $^{6}$University of Toronto, Department of Astronomy \& Astrophysics, 50 St. George Street, Toronto, ON M5S 3H4, Canada\\
  $^{7}$Departamento de Astrofisica, Centro de Astrobiologia (INTA-CSIC), ESAC Campus, PO Box 78, E-28691 Villanueva de la Canada, Spain\\
  $^{8}$Max-Planck Institut f\"ur extraterrestrische Physik, Giessenbachstrasse, D-85741 Garching, Germany\\
  $^{9}$Centro de Astrobiologia (CSIC/INTA), Carretera Ajalvir km 4,  28850 Torrejon de Ardoz, Madrid, Spain\\
  $^{10}$Main Astronomical Observatory, Academy of Sciences of
  Ukraine, Golosiiv Woods, Kyiv-127, 03680 Ukraine}

\begin{document}
\date{11th May 2012}
\pagerange{\pageref{firstpage}--\pageref{lastpage}} \pubyear{2012}
\maketitle
\label{firstpage}

\begin{abstract}
  We report the discovery of $16$ detached M-dwarf eclipsing binaries
  with $J<16$ mag  and provide a detailed characterisation of three
  of them, using high-precision infrared light curves from the WFCAM
  Transit Survey (WTS). Such systems provide the most accurate and
  model-independent method for measuring the fundamental parameters of
  these poorly understood yet numerous stars, which currently lack
  sufficient observations to precisely calibrate stellar evolution
  models. We fully solve for the masses and radii of three of the
  systems, finding orbital periods in the range $1.5<P<4.9$ days, with
  masses spanning $0.35-0.50\msun$ and radii between $0.38-0.50\rsun$,
  with uncertainties of $\sim3.5-6.4\%$ in mass and $\sim2.7-5.5\%$ in
  radius.  Close-companions in short-period binaries are expected to
  be tidally-locked into fast rotational velocities, resulting in high
  levels of magnetic activity. This is predicted to inflate their
  radii by inhibiting convective flow and increasing star spot
  coverage. The radii of the WTS systems are inflated above model
  predictions by $\sim3-12\%$, in agreement with the observed trend,
  despite an expected lower systematic contribution from star spots
  signals at infrared wavelengths. We searched for correlation between
  the orbital period and radius inflation by combining our results
  with all existing M-dwarf radius measurements of comparable
  precision, but we found no statistically significant evidence for a
  decrease in radius inflation for longer period, less active
  systems. Radius inflation continues to exists in non-synchronised
  systems indicating that the problem remains even for very low
  activity M-dwarfs. Resolving this issue is vital not only for
  understanding the most populous stars in the Universe, but also for
  characterising their planetary companions, which hold the best
  prospects for finding Earth-like planets in the traditional
  habitable zone.
\end{abstract}
\begin{keywords}
stars: binaries: eclipsing -- stars: late-type -- stars: fundamental
parameters -- stars: rotation -- stars: magnetic field -- surveys.
\end{keywords}
\section{Introduction}\label{sec:intro}
M-dwarfs ($M_{\star}\lesssim0.6\msun$) constitute more than seventy
per cent of the Galactic stellar population \citep{Hen97} and
consequently, they influence a wide-range of astrophysical phenomena,
from the total baryonic content of the universe, to the shape of the
stellar initial mass function. Furthermore, they are fast becoming a
key player in the hunt for Earth-like planets (e.g. \citealt{Nut08,
  Kopp09,Law11}). The lower masses and smaller radii of M-dwarfs mean
that an Earth-like companion causes a deeper transit and induces a
greater reflex motion in its host than it would do to a solar
analogue, making it comparatively easier to detect Earths in the
traditional habitable zones of cool stars. The inferred properties of
exoplanet companions, such as their density, atmospheric structure and
composition, currently depend on a precise knowledge of the
fundamental properties of the host star, such as its mass, radius,
luminosity and effective temperature at a given age. Yet, to date, no
theoretical model of low-mass stellar evolution can accurately
reproduce all of the observed properties of M-dwarfs
\citep{Hil04,Lop05}, which leaves their planetary companions open to
significant mischaracterisation. Indeed, the characterisation of the
atmosphere of the super-Earth around the M-dwarf GJ 1214 seems to
depend on the spot coverage of the host star \citep{Moo12}.

Detached, double-lined, M-dwarf eclipsing binaries (MEBs) provide the
most accurate and precise, model-independent means of measuring the
fundamental properties of low-mass stars \citep{And91}, and the
coevality of the component stars, coupled with the assumption that
they have the same metallicity due to their shared natal environment,
places stringent observational constraints on stellar evolution
models. In the best cases, the uncertainties on the masses and radii
measured using MEBs can be just $0.5\%$ \citep{Mor09,Kra11a}. However,
since M-dwarfs are intrinsically faint, only a small number of MEBs
have been characterised so far with suitable accuracy to calibrate
low-mass stellar evolution models, and there are even fewer
measurements below $\sim0.35\msun,$ where stellar atmospheres are
thought to transport energy purely by convection \citep{Chab97}.

More worryingly, existing observations show significant discrepancies
with stellar models. The measured radii of M-dwarfs are inflated by
$5-10\%$ compared to model estimates and their effective temperatures
appear too cool by $3-5\%$ (see e.g. \citealt{Lop05, Rib06,Mor10,
  Tor10, Kra11a}). This anomaly has been known for some time but
remains enigmatic. Bizarrely, the two discrepancies compensate each
other in the mass-luminosity plane such that current stellar models
can accurately reproduce the observed mass-luminosity relationship for
M-dwarfs. Two different physical mechanisms have been suggested as the
cause of this apparent radius inflation: i) metallicity
\citep{Ber06,Lop07} and ii) magnetic activity
\citep{Mul01,Rib06,Tor06,Chab07}.

\citet{Ber06} and \citet{Lop07} used interferometrically-measured
radii of single, low-mass stars to look for correlation between
inflation and metallicity. Both studies found evidence that inactive,
single stars with inflated radii corresponded to stars with higher
metallicity, but this did not hold true for active, fast-rotating
single stars and further studies could not confirm the result
\citep{Dem09}. While metallicity may play a role in the scatter of
effective temperatures for a given mass (the effective temperature
depends on the bolometric luminosity which is a function of
metallicity), it seems unlikely that it is the main culprit of radius
inflation.

The magnetic activity hypothesis is steered by the fact that the large
majority of well-characterised MEBs are in short ($<2$ day) orbits.
Such short period systems found in the field (i.e. old systems) are
expected to be tidally-synchronised with circularised orbits
\citep{Zah77}.  The effect of tidal-locking is to increase magnetic
activity and is a notion that is supported by observations of
synchronous, rapid rotation rates in MEBs, a majority of circular
orbits for MEBs, plus X-ray emission and H$\alpha$ emission from at
least one of the components. It is hypothesised that increased
magnetic activity affects the radius of the star in two ways. Firstly,
it can inhibit the convective flow, thus the star must inflate and
cool to maintain hydrostatic equilibrium. \citet{Chab07} modelled this
as a change in the convective mixing length, finding that
a reduced mixing length could account for the
inflated radii of stars in the partially-radiative mass regime, but it
had negligible effect on the predicted radii of stars in the
fully-convective regime. However, \citet{Jac09} showed that the radii
of young, single, active, fully-convective stars in the open cluster
NGC 2516 could be inflated by up to $50\%$, based on radii derived
using photometrically-measured rotation rates and
spectroscopically-measured projected rotational velocities. This
therefore suggests that inhibition of convective flow is not the only
factor responsible for the radius anomaly.

The second consequence of increased magnetic activity is a higher
production of photospheric spots which has a two-fold effect: i) a
loss of radiative efficiency at the surface, causing the star to
inflate and ii) a systematic error in light curve solutions due to a
loss of circular symmetry caused by a polar distribution of spots.
\citet{Mor10} showed that these two effects could account for
$\sim3\%$ and $0-6\%$ of the radius inflation, respectively, with any
any remaining excess ($0-4\%$) produced by inhibition of convective
efficiency. This however is only under certain generalisations, such
as a $30\%$ spot coverage fraction and a concentration of the spot
distribution at the pole. One would perhaps expect the systematic
error induced by star spots to be wavelength dependant, such that
radius measurements obtained at longer wavelength would be closer to
model predictions.

\citet{Kra11a} searched for correlation between the radius anomaly and
the orbital periods of MEBs, to see if the data and the models
converged at longer periods ($\sim3$ days) where the stellar activity
is less aggravated by fast rotation speeds. They found tentative
evidence to suggest that this is the case but it is currently confined
to the realm of small statistics. Not long after their study, the
MEarth project uncovered a 41-day, non-synchronised, non-circularised,
inactive MEB with radius measurements still inflated on average by
$\sim4\%$, despite a detailed attempt to account for spot-induced
systematics \citep{Irw11}. They suggest that either a much larger spot
coverage than the $30\%$ they assumed is required to explain the
inflation, or perhaps that the equation of state for low-mass stars,
despite substantial progress (see review by \citealt{Chab05}), is
still inadequate.

Clearly, a large sample of MEBs with a wide-range of orbital periods
is key to defining the magnetic activity effect and understanding any
further underlying physical issues for modelling the evolution of
low-mass single stars. This in turn will remove many uncertainties in
the properties of exoplanets with M-dwarf host stars. With that in
mind, this paper presents the discovery of many new MEBs to emerge
from the WFCAM Transit Survey, including a full characterisation to
reasonable accuracy for three of the systems using 4-m class
telescopes, despite their relatively faint magnitudes ($i=16.7-17.6$).

In Section~\ref{sec:discovery}, we describe the WFCAM Transit Survey
(WTS) and its observing strategy, and Section~\ref{sec:obs} provides
additional details of the photometric and spectroscopic data we used
to fully characterise three of the MEBs. In
Section~\ref{sec:identify}, we outline how we identified the MEBs
amongst the large catalogue of light curves in the
WTS. Sections~\ref{sec:indices}-\ref{sec:RVs} present our analysis of
all the available follow-up data used to characterise three of the
MEBs including their system effective temperatures, metallicities,
H$\alpha$ emission and surface gravities, via
analysis of low-resolution spectroscopy, their size-ratio and orbital
elements using multi-colour light curves, and their mass ratios using
radial velocities obtained with intermediate-resolution spectra. These
results are combined in Section~\ref{sec:absdim} to determine
individual masses, radii, effective temperatures. We also calculate
their space velocities and assess their membership to the Galactic
thick and thin disks. Lastly, in Section~\ref{sec:discuss}, we discuss
our results in the context of low-mass stellar evolution models and a
mass-radius-period relationship, as suggested by \citet{Kra11a}.

\section{The WFCAM Transit Survey}\label{sec:discovery}
We identified our new MEBs using observations from the WFCAM Transit
Survey (WTS) \citep{Bir11}. The WTS in an on-going photometric
monitoring campaign that operates on the 3.8m United Kingdom Infrared
Telescope (UKIRT) at Mauna Kea, Hawaii. Its primary and complementary
science goals are: i) to provide a stringent observational constraint
on planet formation theories through a statistically meaningful
measure of the occurrence rate of hot Jupiters around low-mass stars
\citep{Kov12}  and ii) to detect a large sample of
eclipsing binaries stars with low-mass primaries and characterise them
to high enough accuracy such that we strongly constrain the stellar
evolution models describing the planet-hosting M-dwarfs found in the
survey.  The WTS contains $\sim6,000$ early to mid M-dwarfs with
$J\leq16$ mag, covering four regions of the sky which
span a total of 6 square degrees.

We combine the large aperture of UKIRT with the Wide-Field Camera
(WFCAM) infrared imaging array to observe in the $J$-band
($1.25\mu$m), near the peak of the spectral energy distribution (SED)
of a cool star.  Our observing strategy takes advantage of a unique
opportunity offered by UKIRT, thanks to the highly efficient
queue-scheduled operational mode of the telescope. Rather than
requesting continuous monitoring, we noted there was room for a
flexible proposal in the queue, which did not require the very best
observing conditions, unlike most of the on-going UKIRT programmes
that require photometric skies with seeing $<1.3\arcsec$
\citep{Lawr07}. The WTS is therefore designed in such a way that there
is always at least one target field visible and it can observe in
mediocre seeing and thin cloud cover.  We chose four target fields to
give us year-round visibility, with each field passing within 15
degrees of zenith.  To select the fields, we combined 2MASS photometry
and the dust extinction maps of \citet{Sch98} to find regions of sky
that maximised the number of dwarf stars and maximised the ratio of
dwarfs to giants \citep{Cru03}, while maintaining $E(B-V)<0.1$. We
stayed relatively close to the galactic plane to increase the number
of early M-dwarfs, but restricted ourselves to $b>5$ degrees to avoid
the worst effects of overcrowding.

The survey began on August 05, 2007, and the eclipsing systems
presented in this paper are all found in just one of the four WTS
fields. The field is centred on $\rm RA=19h$, $\rm Dec=+36d$,
(hereafter, the 19h field), for which the WTS has its most extensive
coverage, with $1145$ data points as of June 16, 2011. Note that this
field is very close to, but does not overlap with, the Kepler field
\citep{Bat06}, but it is promising that recent work showed the giant
contamination in the Kepler field for magnitudes in a comparable range
to our survey was low ($7\pm3\%$ M-giant fraction for $K_{P}>14$),
\citealt{Mann12}.
\section{Observations and Data Reduction}\label{sec:obs}
\subsection{UKIRT/WFCAM $J$-band photometry}\label{sec:photo-wts}
UKIRT and the WFCAM detector provide the survey with a large database
of infrared light curves in which to search for transiting and
eclipsing systems. The WFCAM detector consists of four
$2048\times2048$ $18\mu$m pixel HgCdTe Rockwell Hawaii-II,
non-buttable, infrared arrays that each cover
$13.65\arcmin\times13.65\arcmin$ and are separated by $94\%$ of a chip
width \citep{Cas07}. Each WTS field covers $1.5$ square degrees of
sky, comprising of eight pointings of the WFCAM paw print, exposing
for a 9-point jitter pattern with 10 second exposures at each
position, and tiled to give uniform coverage across the field. It
takes 15 minutes to observe an entire WTS field ($9\times10{\rm
  s}\times8+$overheads), resulting in a cadence of 4 data points per
hour (corresponding to one UKIRT Minimum Schedulable Block). Unless
there are persistently bad sky conditions at Mauna Kea, due to our
relaxed observing constraints the WTS usually observes only at the
beginning of the night, just after twilight in $>1\arcsec$ seeing when
the atmosphere is still cooling and settling.

The 2-D image processing of the WFCAM observations and the generation
of light curves closely follows that of \citet{Irw07} and is explained
in detail in \citet{Kov12}. We refer the avid reader
to these publications for an in-depth discussion of the reduction
techniques but briefly describe it here. For image processing, we use
the automatically reduced images from the Cambridge Astronomical
Survey Unit
pipeline\footnote{http://casu.ast.cam.ac.uk/surveys$-$projects/wfcam/technical},
which is based on the INT wide-field survey pipeline
\citep{Irw01}. This provides the 2-D instrumental signature removal
for infrared arrays including the removal of the dark and reset
anomaly, the flat-field correction using twilight flats, decurtaining
and sky subtraction. We then perform astrometric calibration using
2MASS stars in the field-of-view, resulting in an astrometric accuracy
of $\sim20-50$ mas after correcting for field and differential
distortion\footnote{\scriptsize{http://casu.ast.cam.ac.uk/surveys$-$projects/wfcam/technical/astrometry}}.
For photometric calibration, the detector magnitude zero-point is
derived for each frame using measurements of stars in the 2MASS Point
Source Catalogue that fall within the same frame \citep{Hod09}.

In order to generate a master catalogue of source positions for each
field in the $J$-band filter, we stack 20 frames taken in the best
conditions (i.e. seeing, sky brightness and transparency) and run our
source detection software on the stacked image
\citep{Irw85,Irw01}. The resulting source positions are used to
perform co-located, variable, `soft-edged' (i.e. pro-rata flux division
for boundary pixels) aperture photometry on all of the time-series
images (see \citealt{Irw07}).

For each of the four WFCAM detector chips, we model the flux residuals
in each frame as a function of position using a 2-D quadratic
polynomial, where the residuals are measured for each object as the
difference between its magnitude on the frame in question and its
median magnitude calculated across all frames. By subtracting the
model fit, this frame-to-frame correction can account for effects such
as flat-fielding errors, or varying differential atmospheric
extinction across each frame, which can be significant in wide-field
imaging (see e.g. \citealt{Irw07}).

Our source detection software flags any objects with overlapping
isophotes. We used this information in conjunction with a
morphological image classification flag also generated by the pipeline
to identify non-stellar or blended objects. The plate scale of WFCAM
($0.4\arcsec$/pix) is significantly smaller than those of most small
aperture, ground-based transit survey instruments, such as SuperWASP
\citep{Pol06}, HATNet \citep{Bak04} and TrES \citep{Dun04}, and can
have the advantage of reducing the numbers of blended targets, and
therefore the numbers of transit mimics, despite observing fainter
stars.

The last step in the light curve generation is to perform a correction
for residual seeing-correlated effects caused by image blending that
are not removed by the frame-to-frame correction. For each light
curve, we model the deviations from its median flux as a function of
the stellar image FWHM on the corresponding frame, using a quadratic
polynomial that we then subtract. We note that this method addresses
the symptoms, but not the cause, of the effects of blending.

\begin{figure}
  \centering
  \includegraphics[width=0.5\textwidth]{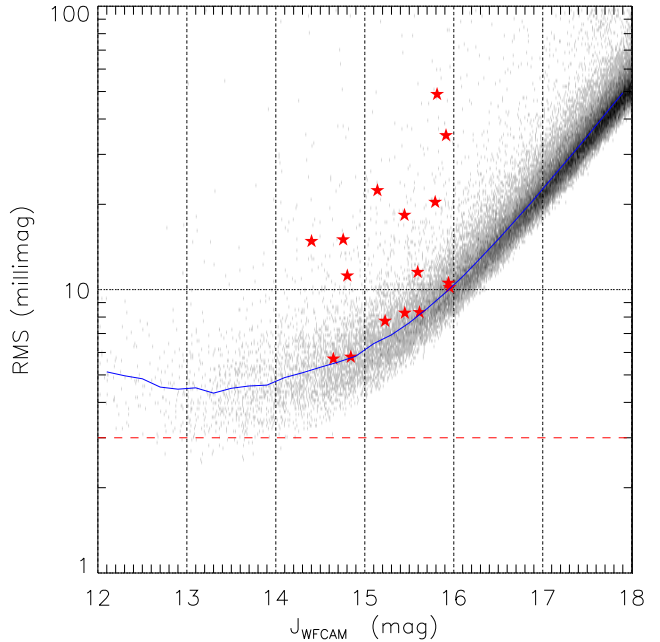}
  \caption{The RMS scatter per data point of the WTS light curves as a
    function of WFCAM $J$ magnitude, for sources in the 19hr field
    with stellar morphological classification. The RMS
      is a robust estimator calculated as $1.48$ $\times$ the median
      of the absolute deviations. We achieve a per data point
    photometric precision of $3-5$ mmag for the brightest objects,
    with a median RMS of $\sim1\%$ for $J=16$
      mag. Saturation occurs between $\sim12-13$ mag
      as it varies across the field and with seeing. The dashed red
      horizontal line at $3$ mmag marks the limit of our photometric
      precision. The blue solid curve shows the median RMS in bins of
      $0.2$ mag. The red stars show the positions of the $16$ WTS
    19hr field MEBs. The shorter period MEBs sit higher in the
    plot. RMS values are given in Table~\ref{tab:others}}
  \label{fig:rms}
\end{figure}

Figure~\ref{fig:rms} shows the per data point photometric precision of
the final light curves for the stellar sources in the 19hr field. The
RMS is calculated as a robust estimator using as $1.48\times${\sc
  mad}, i.e. the equivalent of the Gaussian RMS, where the {\sc mad}
is the median of the absolute deviations \citep{Hoa83}. The upturn
between $J\sim12-13$ mag marks the saturation limit, so for our
brightest objects, we achieve a per data point precision of $\sim3-5$
mmag. The blue solid line shows the median RMS in bins of $0.2$
mag. The median RMS at $J=16$ mag is $\sim1\%$ ($\sim10$ mmag), with a
scatter of $\sim0.8-1.5\%$, and only $5\%$ of sources have an RMS
greater than $15$ mmag at this magnitude. Hence, for the majority of
sources with $J\leq16$ mag, the precision is in theory suitable for
detecting not only M-dwarf eclipsing binaries but also transits of
mid-M stars by planets with radii $\sim1\rearth$ (see \citealt{Kov12}
for the WTS sensitivity to Jupiter- and Neptune-sized planets). The
$16$ new MEBs are shown on the plot by the red star symbols. Note that
shorter period MEBs sit higher on the RMS diagram, but that genuine
longer period MEBs still have RMS values close to the median, due to
our robust estimator and the long observing baseline of the survey.

For the MEB light curves characterised in this paper, we perform an
additional processing step, in which we use visual examination to clip
several clear outlying data points at non-consecutive epochs.

The WTS $J$-band light curve data for the MEBs reported in this paper
are given in Table~\ref{tab:wts_lcs}. We have adopted a naming system
that uniquely identifies each source handled by our data reduction
process, and thus we refer to MEBs characterised in this paper as:
19b-2-01387, 19c-3-01405, and 19e-3-08413. The first number in the
naming strategy gives the Right Ascension hour the target field. The
following letter corresponds to one of the eight pointings that make
up the whole WTS target field.  The number between the hyphens denotes
which of the four WFCAM chips the source is detected on and the final
5 digits constitute the source's unique sequence number in our master
catalogue of WTS sources.

\begin{table*}
  \centering
  \begin{tabular}{@{\extracolsep{\fill}}lccccccc}
    \hline
    \hline
    Name&HJD&$J_{\rm WTS}$&$\sigma_{J_{\rm WTS}}$&$\Delta m_{0}^{a}$&FWHM$^{b}$&$x^{c}$&$y^{c}$\\
    &&(mag)&(mag)&(mag)&(pix)&(pix)&(pix)\\
    \hline
    19b-2-01387&2454317.808241&14.6210&0.0047&0.0001&2.17&321.98&211.07\\
    19b-2-01387&2454317.820311&14.6168&0.0047&0.0002&2.37&321.74&210.88\\
    ...&...&...&...&...&...&...&...\\
    \hline
  \end{tabular}
  \caption{ The WTS $J$-band light curves of 19b-2-01387, 19c-3-01405
    and 19e-3-08413. Magnitudes are given in the WFCAM system.
    \citet{Hod09} provide conversions for other systems. The errors,
    $\sigma_{J}$, are estimated using a standard noise model,
    including contributions from Poisson noise in the stellar counts,
    sky noise, readout noise and errors in the sky background
    estimation. 
    $^{a}$ Correction to the frame magnitude zero point applied in the
    differential photometry procedure. More negative numbers indicate
    greater losses. See \citet{Irw07}. 
    $^{b}$ Median FWHM of the stellar images on the frame. 
    $^{c}$ $x$ and $y$ pixel coordinates the MEB systems on the image,
    derived using a standard intensity-weighted moments
    analysis.
    (This table is published in full in the online journal
    and is shown partially here for guidance regarding its form and content.)}
  \label{tab:wts_lcs}
\end{table*}

Some sources in the WTS fields are observed multiple times during a
full field pointing sequence due to the slight overlap in the exposed
areas in the tile pattern. 19c-3-01405 is one such target, receiving
two measurements for every full field sequence. The median magnitude
for 19c-3-01405 on each chip differs by 32 mmag.  \citet{Hod09} claim
a photometric calibration error of $1.5\%$ for WFCAM thus the median
magnitudes have a $\sim2\sigma$ calibration error. The photometric
calibration uses 2MASS stars that fall on chip in question, so
different calibration stars are used for different chips and
pointing. We combined the light curves from both exposures to create a
single light curve with $893+898=1791$ data points, after first
subtracting the median flux from each light curve. The combined light
curve has the same out-of-eclipse RMS, $8$ mmag, as the two single
light curves. The other two MEBs, 19b-2-01387 and 19e-3-08413, have
$900$ and $899$ data points and an out-of-eclipse RMS of $5$ mmag and
$7$ mmag, respectively.

We also obtained single, deep exposures of each WTS field in the WFCAM
$Z$, $Y$, $J$, $H$ and $K$ filters (exposure times $180, 90, 90,
4\times90$ and $4\times90$ seconds, respectively). These are used in
conjunction with $g, r, i$ and $z$ photometry from SDSS DR7 to create
SEDs and derive first estimates of the effective temperatures for all
sources in the field, as described in Section~\ref{sec:SED}.

\subsection{INT/WFC $i$-band follow-up photometry}
Photometric follow-up observations to help test and refine our light
curve models were obtained in the Sloan $i$-band using the Wide Field
camera (WFC) on the 2.5m Isaac Newton Telescope (INT) at Roque de Los
Muchachos, La Palma. We opted to use the INT's Sloan $i$ filter rather
than the RGO I-band filter as i) it has significantly less fringing
and, ii) unlike the RGO filter, it has a sharp cut-off at $\sim 8500$
\AA~ and therefore avoids strong, time-variable telluric water vapour
absorption lines, which could induce systematics in our time-series
photometry \citep{BaiJ03}. The observing run, between July 18 - August
01, 2010, was part of a wider WTS follow-up campaign to confirm
planetary transit candidates and thus only a few windows were
available to observe eclipses. Using the WFC in fast mode (readout
time 28 sec. for 1$\times$1 binning), we observed a full secondary
eclipse of 19b-2-01387 and both a full primary and a full secondary
eclipse of 19e-3-08413. The observations were centred around the
expected times of primary and secondary eclipse, allowing at least
$30$ minutes of observation either side of ingress and egress to
account for any uncertainty in our predicted eclipse times based on
the modelling of the WTS light curves. In total, we observed $120$
epochs for the secondary eclipse of 19b-2-01387 using $60$s exposures,
and $89$ and $82$ data points for the primary and secondary eclipse of
19e-3-08413, respectively, using $90$s exposures.

We reduced the data using custom built {\sc idl} routines to perform
the standard 2-D image processing (i.e. bias subtraction and
flat-field division). Low-level fringing was removed by subtracting a
scaled super sky-frame. To create the light curves, we performed
variable aperture photometry using circular apertures with the {\sc
  idl} routine {\sc aper}. The sky background was estimated using a
3$\sigma$-clipped median on a 30$\times$30 pixel box,
rejecting bad pixels. For each MEB, we selected sets of 15-20 bright,
nearby, non-saturated, non-blended reference stars to create a master
reference light curve. For each reference star, we selected the
aperture with the smallest out-of-eclipse RMS. We removed the airmass
dependence by fitting a second order polynomial to the out-of-eclipse
data.

The INT $i$-band light curve data is presented in
Table~\ref{tab:int_lc}. The RMS of the out-of-eclipse data for the
primary eclipse of 19b-2-01387 is $4.4$ mmag while the out-of-eclipse
RMS values for the primary and secondary eclipses of 19e-3-08413 are
$5.7$ mmag and $7.1$ mmag, respectively.

\begin{table}
  \begin{center}
    \begin{tabular}{@{\extracolsep{\fill}}cccc}
      \hline
      \hline
      Name&HJD&$\Delta m_{i_{\rm INT}}$&$\sigma_{m_{i_{\rm INT}}}$\\
      &&(mag)&(mag)\\
      \hline
      19b-2-01387&2455400.486275&-0.0044&-0.0034\\
      19b-2-01387&2455400.487652&-0.0049&-0.0024\\
      ...&...&...&...\\
      \hline
    \end{tabular}
    \caption{ INT $i$-band follow-up light curves of 19b-2-01387 and
      19e-3-08413. $\Delta m_{i_{\rm INT}}$ are the differential
      magnitudes with respect to the median of the out-of-eclipse
      measurements such that the out-of-eclipse magnitude is
      $m_{i_{\rm INT}}=0$. The errors, $\sigma_{i}$, are the scaled
      Gaussian equivalents of the median absolute deviation of the
      target from the reference at each epoch
      i.e. $\sigma_{i}\sim1.48\times \rm MAD$.  (This table is
      published in full in the online journal and is shown partially
      here for guidance regarding its form and content.)}
    \label{tab:int_lc}  
  \end{center}
\end{table}

\subsection{IAC80/CAMELOT $g$-band follow-up photometry}
We obtained a single primary eclipse of 19e-3-08413 in the Sloan
$g$-band filter using the CAMELOT CCD imager on the 80cm IAC80
telescope at the Observatorio del Teide in Tenerife. The observations
were obtained on the night of 08 August 2009, during a longer run to
primarily follow-up WTS planet candidates. Exposure times were $60$
seconds and were read out with $1\times1$ binning of
the full detector, resulting in a cadence of $71$ seconds, making a
total of 191 observations for the night.

The time-series photometry was generated using the VAPHOT
package\footnote{http://www.iac.es/galeria/hdeeg/} \citep{Dee01}. The
bias and flat field images were processed using standard {\sc iraf}
routines in order to calibrate the raw science images. The light curve
was then generated using VAPHOT, which is a series of modified {\sc
  iraf} routines that performs aperture photometry; these routines
find the optimum size aperture that maximize the signal-to-noise ratio
for each star. The user can specify whether to use a variable aperture
to account for a time-variable point-spread-function (e.g. due to
changes in the seeing) or to fix it for all images. For this data set,
we fixed the aperture and used an ensemble of $6$ stars with a similar
magnitude to the target to create a master reference light
curve. Finally, a second order polynomial was fitted to the
out-of-eclipse data the target light curve to remove a long-term
systematic trend.

The $g$-band light curve is shown in the bottom left panel of
Figure~\ref{fig:19elc}, and the data are given in
Table~\ref{tab:iac80_lc}. The out-of-eclipse RMS for the target is
$26.9$ mmag, which is higher than the follow-up with the INT, due to
the smaller telescope diameter.

\begin{table}
  \begin{center}
    \begin{tabular}{@{\extracolsep{\fill}}ccc}
      \hline
      \hline
      HJD&$\Delta m_{g_{\rm IAC80}}$&$\sigma_{m_{g_{\rm IAC80}}}$\\
      &(mag)&(mag)\\
      \hline
2455052.51020&-0.0417&0.0290\\
2455052.51113&-0.0091&0.0301\\
      ...&...&...\\
      \hline
    \end{tabular}
    \caption{IAC80 $g$-band follow-up light curve of
      19e-3-08413. $\Delta m_{g_{\rm IAC80}}$ are the differential
      magnitudes with respect to the median of the out-of-eclipse
      measurements such that the out-of-eclipse magnitude is
      $m_{g_{\rm IAC80}}=0$. The errors, $\sigma_{g}$, are those
      computed by the {\sc iraf.phot} package. (This table is
      published in full in the online journal and is shown partially
      here for guidance regarding its form and content.)}
    \label{tab:iac80_lc}  
  \end{center}
\end{table}

\subsection{WHT low-resolution spectroscopy}
We carried out low-resolution spectroscopy during a wider follow-up
campaign of the WTS MEB and planet candidates on several nights
between July 16 and August 17, 2010, using the William Herschel
Telescope (WHT) at Roque de Los Muchachos, La Palma. These spectra
allow the identification of any giant contaminants via gravity
sensitive spectral features, and provide estimates of the
effective system temperatures, plus approximate metallicities and
chromospheric activity indicators (see section~\ref{sec:indices}).

We used the Intermediate dispersion Spectrograph and Imaging System
(ISIS) and the Auxiliary-port Camera (ACAM) on the WHT to obtain our
low-resolution spectra. In all instances we used a $1.0\arcsec$ slit.
We did not use the dichroic during the ISIS observations because it
can induce systematics and up to $10\%$ efficiency losses in the red
arm, which we wanted to avoid given the relative faintness of our
targets. Wavelength and flux calibrations were performed using
periodic observations of standard lamps and spectrophotometric
standard stars throughout the nights. Table~\ref{tab:spectroscopic}
summarises our low-resolution spectroscopic observations.

\begin{table}
  \centering
  \begin{tabular}{@{\extracolsep{\fill}}l@{\hspace{3pt}}r@{\hspace{7pt}}r@{\hspace{7pt}}r@{\hspace{7pt}}r@{\hspace{7pt}}r@{\hspace{7pt}}r@{\hspace{7pt}}}
    \hline
    \hline
    Name&Epoch$^{a}$&$t_{\rm int}$& Instr.&$\lambda_{\rm range}$&R&SNR\\
    &&(s)&&(\AA)&&\\
    \hline
    19b-2-01387&394.71&300&ISIS&6000-9200&$1000$&27\\
    19c-3-01405&426.53&900&ACAM&3300-9100&$450$&30\\
    19e-3-08413&426.54&900&ACAM&3300-9100&$450$&30\\
   \hline
  \end{tabular}
  \caption{Summary of low resolution spectroscopic observations at the
    William Herschel Telescope, La Palma. $^{a}$ JD-2455000.0. }
  \label{tab:spectroscopic}
\end{table}

The reduction of the low-resolution spectra was performed with a
combination of {\sc iraf} routines and custom {\sc idl} procedures. In
{\sc idl}, the spectra were trimmed to encompass the length of the
slit, bias-subtracted and median-filtered to remove cosmic rays. The
ACAM spectra were also flat-fielded. We corrected the flat fields for
dispersion effects using a pixel-integrated sensitivity function. The
{\sc iraf.apall} routine was used to identify the spectra, subtract
the background and optimally sum the flux in apertures along the
trace. For the ISIS spectrum, wavelength and flux calibration was
performed with the CuNe+CuAr standard lamps and ING flux standard
SP2032+248. For ACAM, arc frames were used to determine the wavelength
solution along the slit using a fifth order spline function fit with
an RMS $\rm \sim0.2$\AA. For flux-calibration, we obtained reference
spectra of the ING flux standard SP2157+261.

\subsection{WHT/ISIS intermediate-resolution spectroscopy}\label{sec:hires}
Modelling the individual radial velocities (RVs) of components in a
binary system provides their mass ratio and a lower
  limit on their physical separation. Combining this information with
an inclination angle determined by the light curve of an eclipsing
system ultimately yields the true masses and radii of the stars in the
binary.

We measured the RVs of the components in our MEBs using spectra
obtained with the intermediate-resolution, single-slit spectrograph
ISIS mounted on the WHT. We used the red arm with the R1200R grating
centred on $8500$\AA, giving a wavelength coverage of
$\sim8100-8900$\AA. The slit width was chosen to match the
approximate seeing at the time of observation giving an average
spectral resolution $R\sim9300$. 

The spectra were processed entirely with {\sc iraf}, using the {\sc
  ccdproc} packages for instrumental signature removal. We optimally
extracted the spectra for each object on each night and performed
wavelength and flux calibration using the semi-automatic {\sc
  kpno.doslit} package. Wavelength calibration was achieved using CuNe
arc lamp spectra taken after each set of exposures and flux
calibration was achieved using observations of spectrophotometric
standards.

\subsubsection{Radial velocities via cross-correlation}
The region $8700-8850$\AA ~contains a number of relatively strong
metallic lines present in M-dwarfs and is free of telluric absorption
lines making it amenable for M-dwarf RV measurements \citep{Irw09b}.
We used the {\sc iraf} implementation of the standard 1-D
cross-correlation technique, {\sc fxcor}, to extract the RV
measurements for each MEB component using synthetic spectra from the
MARCS\footnote{http://marcs.astro.uu.se/} spectral database
\citep{Gus08} as templates. The templates had plane-parallel model
geometry, a temperature range from 2800-5500K incremented in 200K
steps, solar metallicity, surface gravity $\log(g)=5.0$ and
a $2$ km/s micro-turbulence velocity, which are all consistent with
low-mass dwarf stars. The best-matching template i.e. the one that
maximised the cross-correlation strength of the primary component for
each object, was used to obtain the final RVs of the system, although
note that the temperature of the best-matching cross-correlation
template is not a reliable estimate of the true effective temperature.
The saturated near-infrared Ca II triplet lines at $8498, 8542$ and
$8662$\AA ~were masked out during the cross-correlation. A summary
of our observations and the extracted radial velocities are given in
table~\ref{tab:hires}.

 \begin{table*}
   \centering
   \begin{tabular}{@{\extracolsep{\fill}}lcrlrrrr}
     \hline
     \hline
     Name&HJD&Slit&$t_{int}$&SNR&Phase&$\rm
     RV_{1}$&$\rm RV_{2}$\\
     &&($\arcsec$)&(n$\times$sec)&&&(km/s)&(km/s)\\
     \hline
     19b-2-01387&     2455395.55200&1.2&$2\times1200$&22.7&0.1422&-143.2&8.0\\
     19b-2-01387&     2455396.46471&0.7&$3\times600$&6.22&0.7513&23.7&-158.0\\
     19b-2-01387&     2455407.52383&1.0&$3\times900$&14.0&0.1314&-137.9&-4.2\\
     19b-2-01387&     2455407.62644&1.0&$3\times1200$&8.0&0.1998&-155.3&25.1\\
     19b-2-01387&     2455408.38324&1.0&$3\times900$&9.1&0.7049&14.5&-157.6\\
     19b-2-01387&     2455408.51689&1.0&$3\times1200$&12.8&0.7941&15.1&-153.7\\
     19b-2-01387&     2455408.63070&1.0&$3\times1200$&13.4&0.8700&-9.8&-139.2\\
     19b-2-01387&     2455409.38673&1.0&$3\times1200$&14.3&0.3745&-128.4&-4.8\\
     \hline
     19c-3-01405&     2455407.43073&1.0&$1200+630$&6.4&0.2244&-62.5&57.0\\
     19c-3-01405&     2455407.47937&1.0&$3\times1200$&5.3&0.2343&-57.0&52.7\\
     19c-3-01405&     2455407.58012&1.0&$3\times1200$&5.3&0.2547&-63.8&54.6\\
     19c-3-01405&     2455408.46929&1.0&$3\times1200$&6.0&0.4347&-21.7&22.0\\
     19c-3-01405&     2455409.56881&1.0&$3\times1200$&6.0&0.6573&47.3&-52.6\\
     19c-3-01405&     2455409.68190&0.8&$3\times1200$&5.1&0.6802&42.5&-64.4\\
     19c-3-01405&     2455409.47707&0.8&$3\times1200$&7.5&0.6387&46.4&-43.6\\
     \hline
     19e-3-08413&     2455408.42993&1.0&$3\times1200$&7.1&0.6640&108.0&-46.5\\
     19e-3-08413&     2455408.56307&1.0&$3\times1200$&8.7&0.7435&113.1&-58.4\\
     19e-3-08413&     2455409.43629&1.0&$3\times1200$&8.9&0.2654&-24.8&140.9\\
     19e-3-08413&     2455409.52287&0.8&$3\times1200$&7.5&0.3171&-27.6&125.6\\
     19e-3-08413&     2455409.61343&0.8&$3\times1200$&7.5&0.3712&-9.4&109.1\\
     \hline
\end{tabular}
\caption{Summary of intermediate-resolution spectroscopic
  observations. All observations were centred on $8500$\AA.}
\label{tab:hires}
\end{table*}

\section{Identification of M-dwarf Eclipsing Binaries}\label{sec:identify}
\subsection{The M-dwarf sample}\label{sec:SED}
\begin{figure}
  \centering
  \includegraphics[width=0.49\textwidth]{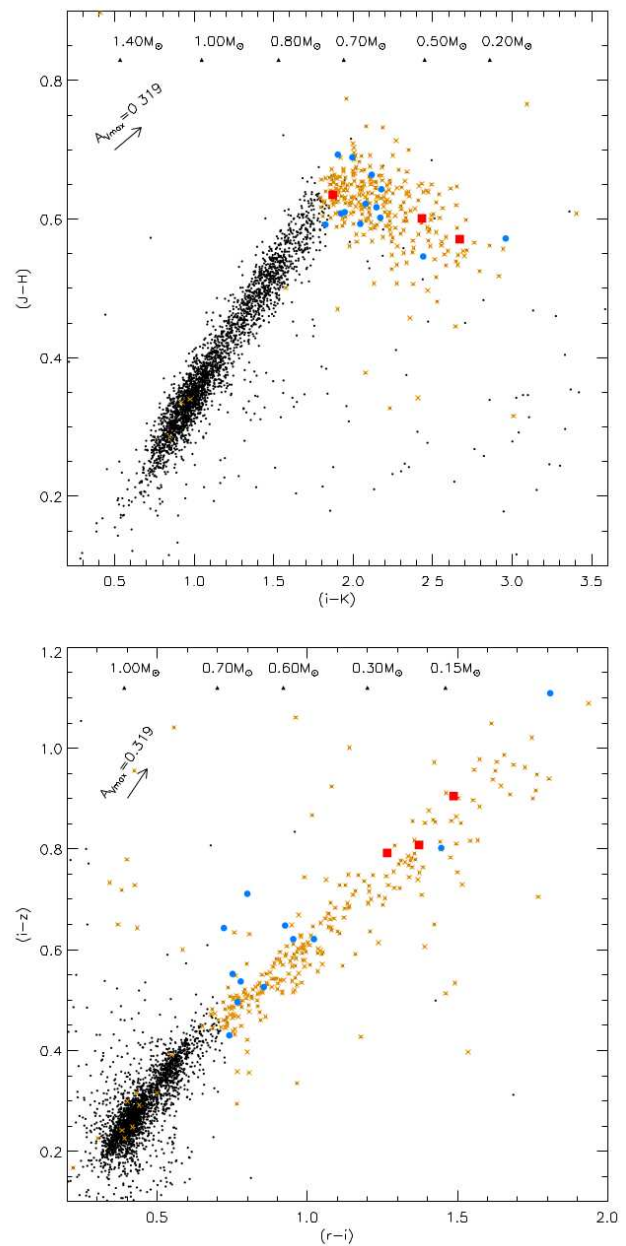}
  \caption{Colour-colour plots of the sources in one of the WFCAM
    pointings for the 19hr field (black $+$),
    overlaid with the full 19hr field sample of detached MEB
    candidates (blue filled circles and red filled squares). The
    filled red squares mark the three MEB systems characterised in
    this paper. The orange crosses mark the M-dwarf candidate sources
    in the pointing (see Section~\ref{sec:SED}). The triangles mark
    the masses for the given colour index, derived from the 1 Gyr
    solar metallicity isochrone of the \citet{Bar98} low-mass stellar
    evolution models. The arrows mark the maximum reddening vector,
    assuming a distance of $1$ kpc.}
  \label{fig:cc_wts}
\end{figure}

It is possible to select M-dwarfs in WTS fields using simple
colour-colour plots such as those shown in Figure~\ref{fig:cc_wts},
which were compiled using our deep WFCAM photometry plus magnitudes
from SDSS DR7, which has a fortuitous overlap with the 19hr field.
\citet{Jon94} showed that the $(i-K)$ colour is a reasonable estimator
for the effective temperature, however the eclipsing nature of the
systems we are interested in can cause irregularities in the colour
indices, especially since the WFCAM photometry was taken at different
epochs to each other and the SDSS photometry. For example, a system of
two equal mass stars in total eclipse result is 0.75 mag fainter
compared to its out-of-eclipse magnitude. We made a more robust sample
of M-dwarfs by estimating the effective temperature of each source in
the 19h field via SED fitting of all the available passbands i.e. SDSS
$g, r, i$ and $z$-band plus WFCAM $Z, Y, J, H$ and $K$-band. By
rejecting the most outlying magnitudes from the best SED fit, one
becomes less susceptible to errors from in-eclipse observations. Note
that the SDSS $u$-band magnitudes of our redder sources are affected
by the known red leak in the $u$ filter and are hence excluded from
the SED fitting process.

To perform the SED fitting, we first put all the observed photometry
to the Vega system (see \citealt{Hew06} and \citealt{Hod09} for
conversions). Although the WFCAM photometry is calibrated to $1.5-2\%$
with respect to 2MASS \citep{Hod09}, the 2MASS photometry also carries
its own systematic error, so we assumed an extra $3\%$ systematic
error added in quadrature to the photometric errors for each source to
account for calibration errors between different surveys. We used a
simple $\chi^{2}$ fitting routine to compare the data to a set of
solar metallicity model magnitudes at an age of 1 Gyr from the stellar
evolution models of \citet{Bar98}. We linearly interpolated the model
magnitudes onto a regular grid of $5$ K intervals from $1739-6554$ K,
to enable a more precise location of the $\chi^{2}$ minimum. If the
worst fitting data point in the best $\chi^{2}$ fit was more than a
$5\sigma$ outlier, we excluded that data point and re-ran the fitting
procedure. This makes the process more robust to exposures taken in
eclipse. The errors on the effective temperatures include the formal
$1\sigma$ statistical errors from the $\chi^{2}$ fit (which are likely
to be under-estimated) plus an assumed $\pm100$ K systematic
uncertainty. This error also takes into account the known missing
opacity issue in the optical bandpasses in the \citet{Bar98} models.

Our M-dwarf sample is conservative. It contains any source with an SED
effective temperature $\leq4209$ K, magnitude $J\leq16$
mag and a stellar class morphology flag (as
determined by the data reduction pipeline).  The maximum effective
temperature corresponds to a radius of $0.66\rsun$ at the typical
field star age of 1 Gyr, according to the stellar evolution models of
\citet{Bar98}. We opted to restrict our MEB search to $J\leq16$
mag because the prospects for ground-based radial
velocity follow-up are bleak beyond $J=16$ mag
($I\sim18$ mag, \citealt{Aig07}) if we wish to
achieve accurate masses and radii that provide useful constraints on
stellar evolution models. We found a total of $2,705$ M-dwarf sources
in the 19hr field.

Table~\ref{tab:photometric} gives the single epoch, deep photometry
from SDSS and WFCAM, plus the proper motions from the SDSS DR7
database \citep{Mun04,Mun08} for the systems characterised in this
paper. Their SED-derived system effective temperatures, $\rm T_{\rm
  eff,SED}$ are given in Table~\ref{tab:spec_ind}.

\begin{table}
  \centering
  \begin{tabular}{@{\extracolsep{\fill}}l@{\hspace{3pt}}rrr}
    \hline
    \hline
    Parameter&19b-2-01387&19c-3-01405&19e-3-08413\\
    \hline
    $\alpha_{J2000}$&19:34:15.5&19:36:40.7&19:32:43.2\\
    $\delta_{J2000}$&36:28:27.3&36:42:46.0&36:36:53.5\\
    $\mu_{\alpha}cos\delta$
    ($\arcsec$/yr)&$0.023\pm0.003$&$-0.002\pm0.004$
    &$0.008\pm0.004$\\
    $\mu_{\delta}$ ($\arcsec$/yr)&$0.032\pm0.003$&$-0.001\pm0.004$
    &$-0.007\pm0.004$\\
    $g$&$19.088\pm0.010$&$20.342\pm0.024$&$20.198\pm0.020$\\
    $r$&$17.697\pm0.006$&$18.901\pm0.012$&$18.640\pm0.009$\\
    $i$&$16.651\pm0.004$&$17.634\pm0.008$&$17.488\pm0.005$\\
    $z$&$16.026\pm0.007$&$16.896\pm0.012$&$16.847\pm0.010$\\
    $Z$&$15.593\pm0.005$&$16.589\pm0.007$&$16.156\pm0.006$\\
    $Y$&$15.188\pm0.006$&$16.432\pm0.011$&$15.832\pm0.008$\\
    $J$&$14.721\pm0.004$&$15.706\pm0.006$&$15.268\pm0.005$\\
    $H$&$14.086\pm0.003$&$15.105\pm0.006$&$14.697\pm0.005$\\
    $K$&$14.414\pm0.006$&$14.836\pm0.007$&$14.452\pm0.006$\\
    \hline
  \end{tabular}
  \caption{A summary of photometric properties for the three MEBs, including
    our photometrically derived effective temperatures and spectral types. The
    proper motions $\mu_{\alpha}cos\delta$ and $\mu_{\delta}$
    are taken from the SDSS DR7 database. SDSS magnitudes $g, r, i$
    and $z$ are in AB magnitudes, while the WFCAM $Z, Y, J, H$ and
    $K$ magnitudes are given in the Vega system. The errors on the
    photometry are the photon-counting errors and do not include the
    extra $3\%$ systematic error we add in quadrature when performing
    the SED-fitting. Conversions of the WFCAM magnitudes to other systems can be found in
    \citet{Hod09}. Note that the WFCAM $K$-band magnitude for
    19b-2-01387 was obtained during an eclipse event and does not
    represent the total system magnitude.}
  \label{tab:photometric}
\end{table}

\subsubsection{Interstellar reddening}\label{sec:redden}
The photometry for the 19hr field is not dereddened before performing
the SED fitting. The faint magnitudes of our M-dwarf sources implies
they are at non-negligible distances and that extinction along the
line-of-sight may be significant. This means that our M-dwarf sample
may contain hotter sources than we expect. At $J\le16$
mag, assuming no reddening, the WTS is
distance-limited to $\sim1$ kpc for the earliest M-dwarfs ($M_{J}=6$
mag at 1 Gyr for M0V, $M_{\star}=0.6\msun$, using the models of
\citealt{Bar98}). We investigated the reddening effect in the
direction of the 19hr field using a model for interstellar extinction
presented by \citet{Dri03}. In this model, extinction does not have a
simple linear dependency on distance but is instead a
three-dimensional description of the Galaxy, consisting of a dust
disk, spiral arms as mapped by HII regions, plus a local Orion-Cygnus
arm segment, where dust parameters are constrained by COBE/DIRBE far
infrared observations. Using this model, we calculate that
$A_{V}=0.319$ mag ($E(B-V)=0.103$ mag) at $1$ kpc in the direction of
the 19hr field. We used the conversion factors in Table 6 of
\citet{Sch98} to calculate the absorption in the UKIRT and SDSS
bandpasses, finding $A_{g}=0.370$ mag, $A_{K}=0.036$ mag,
$E(r-i)=0.065$, $E(i-z)=0.059$ and $E(J-H)=0.032$.  The reddening
affect along the line-of-sight to the field thus appears to be
small. We show this maximum reddening vector as an arrow in
Figure~\ref{fig:cc_wts}.

For the most interesting targets in the WTS (EBs or planet
candidates), we obtain low-resolution spectra to further characterise
the systems and check their dwarf-like nature (see
Section~\ref{sec:indices}). Effective temperatures based on spectral
analysis suffer less from the effects of reddening effects because the
analysis depends not only on the slope of the continuum but also the
shape of specific molecular features, unlike the SED fitting.
Therefore, the SED effective temperatures are only a first estimate
and we will later adopt values derived by fitting model atmospheres to
low-resolution spectra of our MEBs (see Section~\ref{sec:spt-teff}).

\subsection{Eclipse detection}\label{sec:detect}
We made the initial detection of our MEBs during an automated search
for transiting planets in the WTS light curves, for which we used the
Box-Least-Squares (BLS) algorithm, {\sc occfit}, as described in
\citet{Aig04}, and employed by \citet{Mil08}. The box represents a
periodic decrease in the mean flux of the star over a short time scale
(an upside-down top hat). The in-occultation data points in the light
curves fall into a single bin, $I$, while the out-of-occultation data
points form the ensemble $O$. This single bin approach may seem
simplistic but in the absence of significant intrinsic stellar
variability, such as star spot modulation, it becomes a valid
approximation to an eclipse and is sufficient for the purpose of
\emph{detection}. Given the relatively weak signal induced by star
spot activity in the $J$-band, we did not filter the light curves for
stellar variability before executing the detection algorithm. We ran
{\sc occfit} on the M-dwarf sample light curves in the 19h field. Our
data invariably suffer from correlated `red' noise, thus we adjust the
{\sc occfit} detection statistic, $S$, which assesses the significance
of our detections, with the procedure described by \citet{Pon06b} to
derive a new statistic, $\sred$. This process is explained in detail
for {\sc occfit} detections in \citet{Mil08}.

\subsection{Candidate selection}
To automatically extract the MEB candidates from results of running
{\sc occfit} on the M-dwarf sample light curves, we required that
$\sred \ge 5$ and that the detected orbital period must not be near
the common window-function alias at one day i.e. $0.99>P>1.005$ days.
This gave $561$ light curves to eyeball, during which we removed
objects with spurious eclipse-like features associated with light
curves near the saturation limit.

In total, we found $26$ sources showing significant eclipse-features
in the 19h field, of which $16$ appear to be detached and have
full-phase coverage, with well-sampled primary and secondary eclipses.
The detached MEB candidates are marked on the colour-colour plot in
Figure~\ref{fig:cc_wts} by the blue filled circles and red filled
squares. The orbital periods of the MEBs corresponding to the blue
filled circles are given in Table~\ref{tab:others} and their folded
light curves are shown in Figures~\ref{fig:others}
and~\ref{fig:others2}. The MEBs corresponding to the red filled
squares are the subjects of the remaining detailed analysis in this
paper.

\section{Low-resolution spectroscopic analysis}\label{sec:indices}
\begin{figure}
  \centering
  \includegraphics[width=0.5\textwidth]{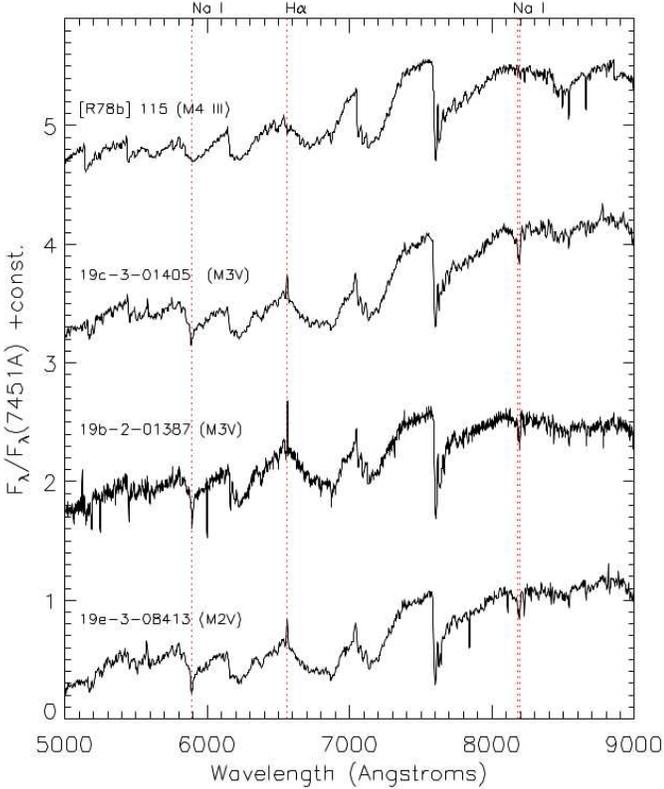}
  \caption{Low-resolution spectra of our three new MEBs plus a known
    M-giant star (top spectrum) for comparison. The TiO absorption
    band at $7100$\AA ~signifies the onset of the M-dwarf spectral
    types. The dotted vertical lines, from left the right, mark the
    Na I, $\rm H\alpha$ and the Na I doublet rest wavelengths in
    air. The Na I doublet is strong in dwarfs while the Calcium
    infrared triplet at $8498, 8542$ and $8662$\AA ~is strongest in
    giants. The deep features at $7594$ and $7685$\AA~
    are telluric $O_{2}$ absorption. $\rm H\alpha$ emission is clearly
    present in all three MEBs.}
  \label{fig:lowres}
\end{figure}

Low-resolution spectra of our three characterised MEBs, as shown in
Figure~\ref{fig:lowres}, permit a further analysis of their composite
system properties and provide consistency checks on the main-sequence
dwarf nature of the systems.

\subsection{Surface Gravity}
\citet{Sle06} and \citet{Lod11} have shown that the depths of alkaline
absorption lines between $6300-8825$\AA~ can highlight low surface
gravity features in low-mass stars. We used the spectral indices $\rm
Na_{8189}$ and $\rm TiO_{7140}$ to search for any giant star
contaminants in the MEBs and found that all three MEBs have indices
consistent with dwarf star gravity.  We note that our low-resolution
spectra were not corrected for telluric absorption, which is prevalent
in the $\rm Na_{8189}$ region, and thus our measured indices may not
be completely reliable. However a visual inspection of the spectra
also reveals deep, clear absorption by the NaI doublet at $8183$\AA,
$8195$\AA~ as highlighted in Figure~\ref{fig:lowres}, which is not
seen in giant stars. For comparison, we also observed an M4III giant
standard star, [R78b] 115, shown at the top of
Figure~\ref{fig:lowres}, with the same set up on the same night. It
lacks the deep Na I doublet absorption lines found in dwarfs and its
measured spectral indices are $\rm TiO_{7140}=2.0\pm0.2$ and $\rm
Na_{8189}=0.97\pm0.04$, which places it in the low-surface gravity
region for M4 spectral types in Figure 11 of \citet{Sle06}. The
gravity-sensitive spectral index values for our MEBs are given in
Table~\ref{tab:spec_ind}.

\subsection{Metallicity}\label{sec:metals}
The profusion of broad molecular lines in M-dwarf spectra, caused by
absorbing compounds such as Titanium Oxide and Vanadium Oxide redwards
of $6000$\AA ~\citep{Kir91}, make it difficult to accurately define
the continuum level, which complicates M-dwarf metallicity
measurements. However, recent work shows that the relative strengths
of metal hydride and metal oxide molecular bands in low-resolution
optical wavelengths can be used to separate metal-poor subdwarfs from
solar-metallicity systems. For example, \citet{Woo09} provided a set
of equal metallicity contours in the plane of the CaH2+CaH3 and TiO5
spectral indices defined by \citet{Rei95}, and they mapped the
metallicity index $\zeta_{\rm TiO/CaH}$ described by \citet{Lep07}
onto an absolute metallicity scale, calibrated by
metallicity measurements from well-defined FGK stars with M-dwarf
companions, albeit with a significant scatter of $\sim0.3$
dex. \citet{Dhi11} have refined the coefficients for $\zeta_{\rm
  TiO/CaH}$ after finding a slight bias for higher metallicity in
early M-dwarfs. We measured the CaH2+CaH3 and TiO5 indices in our MEB
spectra and compared them with these works. We found that all three of
our systems are consistent with solar metallicity. The measured values
of the metallicity-sensitive indices for our MEBs are given in
Table~\ref{tab:spec_ind}.

One should note that further progress has been made in M-dwarf
metallicity measurements by moving to the infrared and using both
low-resolution $K$-band spectra \citep{Roj10,Mui11} and
high-resolution $J$-band spectra \citep{One11,delBur11}. These regions are
relatively free of molecular lines, allowing one to isolate atomic
lines (such as Na I and Ca I) and thus achieve a precise continuum
placement. However, in the spectra of M-dwarf short period binary
systems, one must be aware that the presence of double-lines and
rotationally-broadened features further increase the uncertainty in
their metallicity estimates.

\subsection{H$\alpha$ Emission}
All three of our MEBs show clear $\rm H\alpha$ emission in their
low-resolution spectra, although it is not possible to discern if both
components are in emission. The equivalent widths of these lines,
which are a measure of the chromospheric activity, are reported in
Table~\ref{tab:spec_ind}, where a negative symbol denotes emission.
H$\alpha$ emission can be a sign of youth, but we do not see any
accompanying low-surface gravity features. The strength of the
H$\alpha$ emission seen in our MEBs is comparable with other close
binary systems (e.g. \citealt{Kra11a}) and thus is most likely caused
by high magnetic activity in the systems. None of the systems have
equivalent widths $< -8$\AA, which places them in the non-active
accretion region of the empirically derived accretion criterion of
\citet{Barr03}.

\subsection{Spectral type and effective temperature}\label{sec:spt-teff}
Our low-resolution spectra permit an independent estimate of the
spectral types and effective temperatures of the MEBs to compare with
the SED fitting values. Initially, we assessed the spectral types
using the {\sc
  Hammer}\footnote{http://www.astro.cornell.edu/$\sim$kcovey/thehammer.html}
spectral-typing tool, which estimates MK spectral types by measuring a
set of atomic and molecular features \citep{Cov07}. One can visually
inspect the automatic fit by eye and adjust the fit interactively. For
the latest-type stars (K and M), the automated characterisation is
expected to have an uncertainty of $\sim2$ subclasses. We found that
19b-2-01387 has a visual best-match with an M2V system, while the
other two MEBs were visually closest to M3V systems. M-dwarf studies
\citep{Rei95,Giz97} have found that the TiO5 spectral index could also
be used to estimate spectral types to an accuracy of $\pm0.5$
subclasses for stars in the range K7V-M6.5V. The value of this index
and the associated spectral type (SpT) are given for each of our three
MEBs in Table~\ref{tab:spec_ind}. We find a reasonable agreement
between the spectral index results, the visual estimates and the SED
derived spectral types.

\citet{Woo06} derived a relationship between the CaH2 index and the
effective temperatures of M-dwarfs in the range $3500 $K$ <\rm T_{\rm
  eff}<4000$ K. Table~\ref{tab:spec_ind} gives the value of this index
and the associated effective temperatures, labelled $\rm T_{\rm eff}$
(CaH2), for our three MEBs. \citet{Woo06} do not quote an uncertainty
on the relationship, so we assumed errors of $\pm150$ K. Within the
assumed errors, the effective temperatures derived from the spectral
indices and the SED fitting agree, but the relationship between the
CaH2 index and $\rm T_{\rm eff}$ has not been robustly tested for the
CaH2 values we have measured.

Instead, we have determined the system effective temperatures for our
MEBs by directly comparing the observed spectra to cool star model
atmospheres using a $\chi^{2}$-minimisation algorithm. This
incorporated the observational errors, which were taken from the error
spectrum produced during the optimal extraction of the spectra. We
used a grid of NextGen atmospheric models \citep{All97} interpolated
to the same resolution as our low-resolution spectra. The models had
increments of $\Delta\teff=100$K, solar metallicity and a surface
gravity $\log(g)=5.0$ (a typical value for early-type field
M-dwarfs), and spanned $5000-8500$\AA. During the fitting, we
masked out the strong telluric $\rm O_{2}$ features at $7594,
7685$\AA~ and the H$\alpha$ emission line at $6563$\AA ~as these
are not present in the models, although we found that their inclusion
had a negligible affect on the results. All the spectra were
normalised to their continuum before fitting. We fitted the
$\chi^2$-distribution for each MEB with a six-order polynomial to
locate its minimum. The corresponding best-fitting $\teff$ (atmos.,
adopted) is given in Table~\ref{tab:spec_ind}. Assuming systematic
correlation between adjacent pixels in the observed spectrum, we
multiplied the formal $1\sigma$ errors from the $\chi^{2}$-fit by
$\sqrt{3}$ to obtain the final errors on the system effective
temperatures.

From here on, our analysis is performed with system effective
temperatures derived from model atmosphere fitting. Although our
different methods agree within their errors, the model atmosphere
fitting is more robust against reddening effects, even if this effect
is expected to be small, as discussed earlier.

\begin{table*}
  \centering
  \begin{tabular}{@{\extracolsep{\fill}} lrrrrrrrrrr}
    \hline
    \hline
    Name&$\teff$&$\teff$&$\teff$&SpT&$\rm TiO5$&$\rm CaH2$&$\rm CaH3$&$\rm TiO_{7140}$&$\rm Na_{8189}$&EW(H$\alpha$)\\
    &(SED)& (atmos., adopted)& (CaH2)& (TiO5)&&&&&&(\AA)\\
    \hline
    19b-2-01387&$3494\pm116$&$3590\pm100$&$3586\pm150$&M$2.7\pm0.5$&
    $0.52$&$0.52$&$0.73$&$1.46$&$0.89$&$-3.2$\\
    19c-3-01405&$3389\pm110$&$3307\pm130$&$3514\pm150$&M$2.8\pm0.5$&
    $0.50$&$0.48$&$0.75$&$1.60$&$0.87$&$-4.3$\\
    19e-3-08413&$3349\pm111$&$3456\pm140$&$3569\pm150$&M$2.3\pm0.5$&
    $0.54$&$0.51$&$0.73$&$1.46$&$0.90$&$-4.1$\\
    \hline
  \end{tabular}
  \caption{A summary of the spectral indices, derived effective 
    temperatures and spectral types (SpT) for the three characterised
    MEBs. The photometric estimates are labelled with (SED). They have
    the smallest errors, which include the formal uncertainties plus a
    $100$ K systematic uncertainty, but they potentially suffer from
    reddening effects
    and under-estimation of the errors. Our adopted effective
    temperatures are marked (atmos., adopted). They are derived
    from comparison with the NextGen model atmosphere spectra \citep{All97}
    and are more robust against reddening effects. The (TiO5) and
    (CaH2) labels mark values derived from the spectral index
    relations of \citet{Rei95} and \citet{Woo06}, respectively. We use
    $\rm T_{\rm eff}$ (atmos., adopted) for all subsequent analysis
    in this paper.}
  \label{tab:spec_ind}
\end{table*}

\section{Light curve analysis}\label{sec:lcanalysis}
Light curves of an eclipsing binary provide a wealth of information
about the system, including its orbital geometry, ephemeris, and the
relative size and relative radiative properties of the stars. We used
the eclipsing binary software, {\sc
  jktEBOP}\footnote{http://www.astro.keele.ac.uk/$\sim$jkt/}
\citep{Sou04b,Sou04c}, to model the light curves of our MEBs. {\sc
  jktEBOP} is a modified version of {\sc EBOP} (Eclipsing Binary Orbit
Program; \citealt{Nel72,Pop81,Etz80}). The algorithm is only valid for
well-detached eclipsing binaries with small tidal distortions, i.e
near-spherical stars with oblateness $<0.04$ \citep{Pop81}. A first
pass fit with {\sc jktEBOP} showed that this criterion is satisfied by
all three of our MEBs. 

The light curve model of a detached, circularised eclipsing binary is
largely independent of its radial velocity model, which allowed us to
perform light curve modelling and derive precise orbital periods on
which to base our follow-up multi-wavelength photometry and radial velocity
measurements. The RV-dependent part of the light curve model is the
mass ratio, $q$, which controls the deformation of the stars. In our
initial analysis to determine precise orbital periods, we assumed
circular stars, which is reasonable for detached systems, but the
observed mass ratios (see Section~\ref{sec:RVs}) were adopted in the
final light curve analysis.

{\sc jktEBOP} depends on a number of physical parameters. We allowed
the following parameters to vary for all three systems during the
final fitting process: i) the sum of the radii as a fraction of their
orbital separation, $(R_{1}+R_{2})/a$, where $R_{j}$ is the stellar
radius and $a$ is the semi-major axis, ii) the ratio of the radii,
$k=R_{2}/R_{1}$, iii) the orbital inclination, $i$, iv) the central
surface brightness ratio, $J$, which is essentially equal to the ratio
of the primary and secondary eclipse depths, v) a light curve
normalisation factor, corresponding to the magnitude at quadrature
phase, vi) $e\rm cos\omega$, where $e$ is the eccentricity and
$\omega$ is the longitude of periastron, vii) $e\rm sin\omega$, viii)
the orbital period, $P$ and ix) the orbital phase zero-point, $T_{0}$,
corresponding to the time of mid-primary eclipse. The starting values
of $P$ and $T_{0}$ are taken from the original {\sc occfit} detection
(see Section~\ref{sec:detect}). In the final fit, the observed $q$
value is held fixed. The reflection coefficients were not fitted,
instead they were calculated from the geometry of the system. The
small effect of gravity darkening was determined by fixing the gravity
darkening coefficients to suitable values for stars with convective
envelopes ($\beta=0.32$) \citep{Luc67}. {\sc jktEBOP}
  will allow for a source of third light in the model, whether it be
  from a genuine bound object or from some foreground or background
  contamination, so we initially allowed the third light parameter to
vary but found it to be negligible in all cases and thus fixed it to
zero in the final analysis.

Our light curves, like many others, are not of sufficient quality to
fit for limb darkening, so we fixed the limb darkening coefficients
for each component star. {\sc jktld} is a subroutine of {\sc jktEBOP}
that gives appropriate limb darkening law coefficients for a given
bandpass based on a database of coefficients calculated from available
stellar model atmospheres. We used the PHOENIX model atmospheres
\citep{Cla00,Cla04} and the square-root limb darkening law in all
cases. Studies such as \citet{VH93} have shown that the square-root
law is the most accurate at infrared wavelengths. For each star, we
assumed surface gravities of $\log(g)=5$, a solar metallicity and
micro-turbulence of $2$ km/s, and used estimated effective
temperatures for the component stars: $[T_{\rm eff,1},T_{\rm
  eff,2}]=$[3500K, 3450K] for 19b-2-01387, $[T_{\rm eff,1},T_{\rm
  eff,2}]=$[3300K, 3300K] for 19e-3-08413, and $[T_{\rm eff,1},T_{\rm
  eff,2}]=$[3525K, 3350K] for 19c-3-01405.

Note that we did not iterate the limb darkening coefficients with the
final derived values of $T_{1}$ and $T_{2}$ (see
Section~\ref{sec:absdim}) as they only differed by $\sim30$ K
($<1\sigma$) from the assumed values.  This would be computationally
intensive to do and would result in a negligible
effect on the final result.

\begin{figure*}
  \centering
  \includegraphics[width=0.49\textwidth,clip=true, trim=0cm -0.5cm 0cm 0cm]{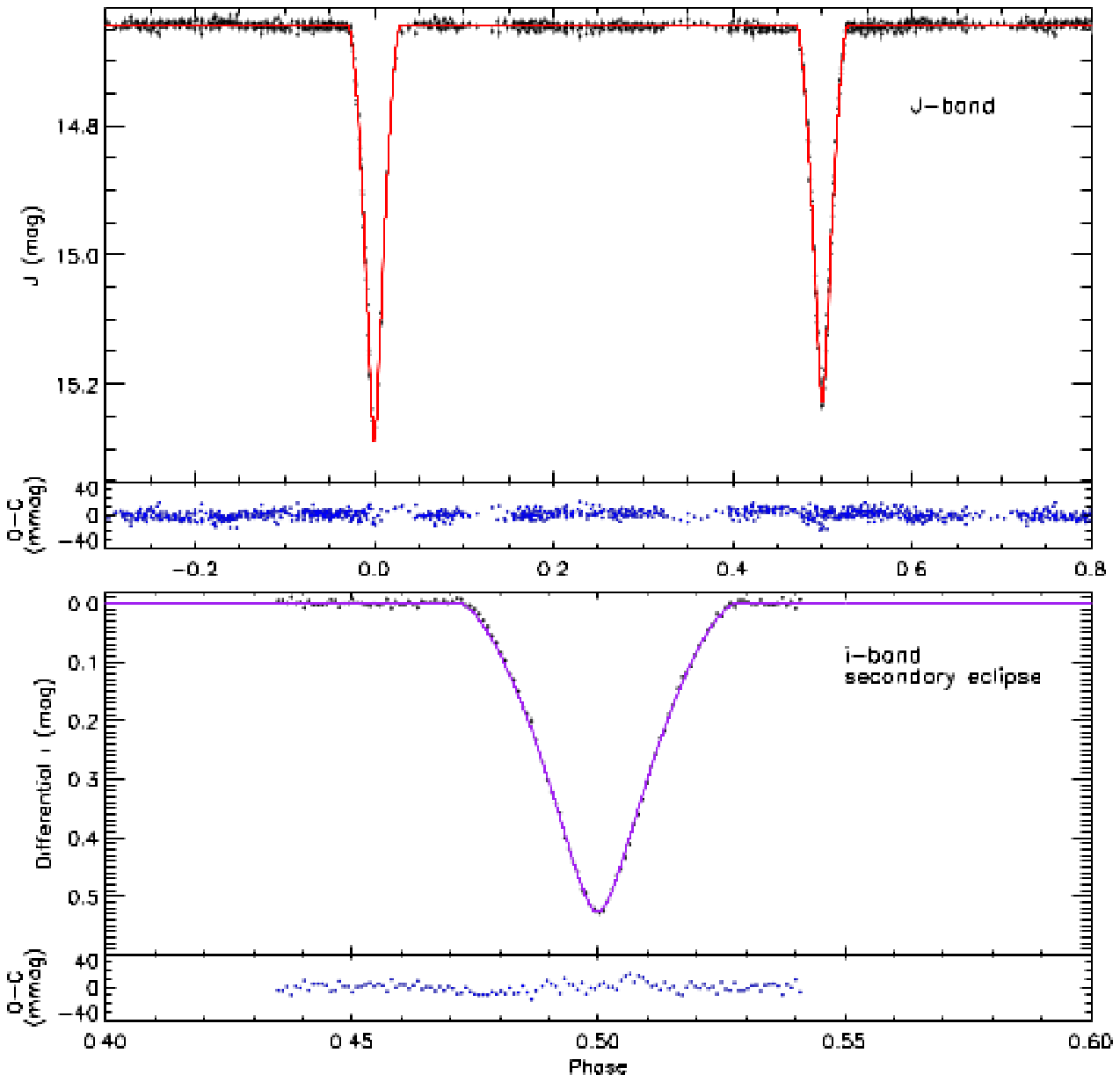}
  \includegraphics[width=0.49\textwidth]{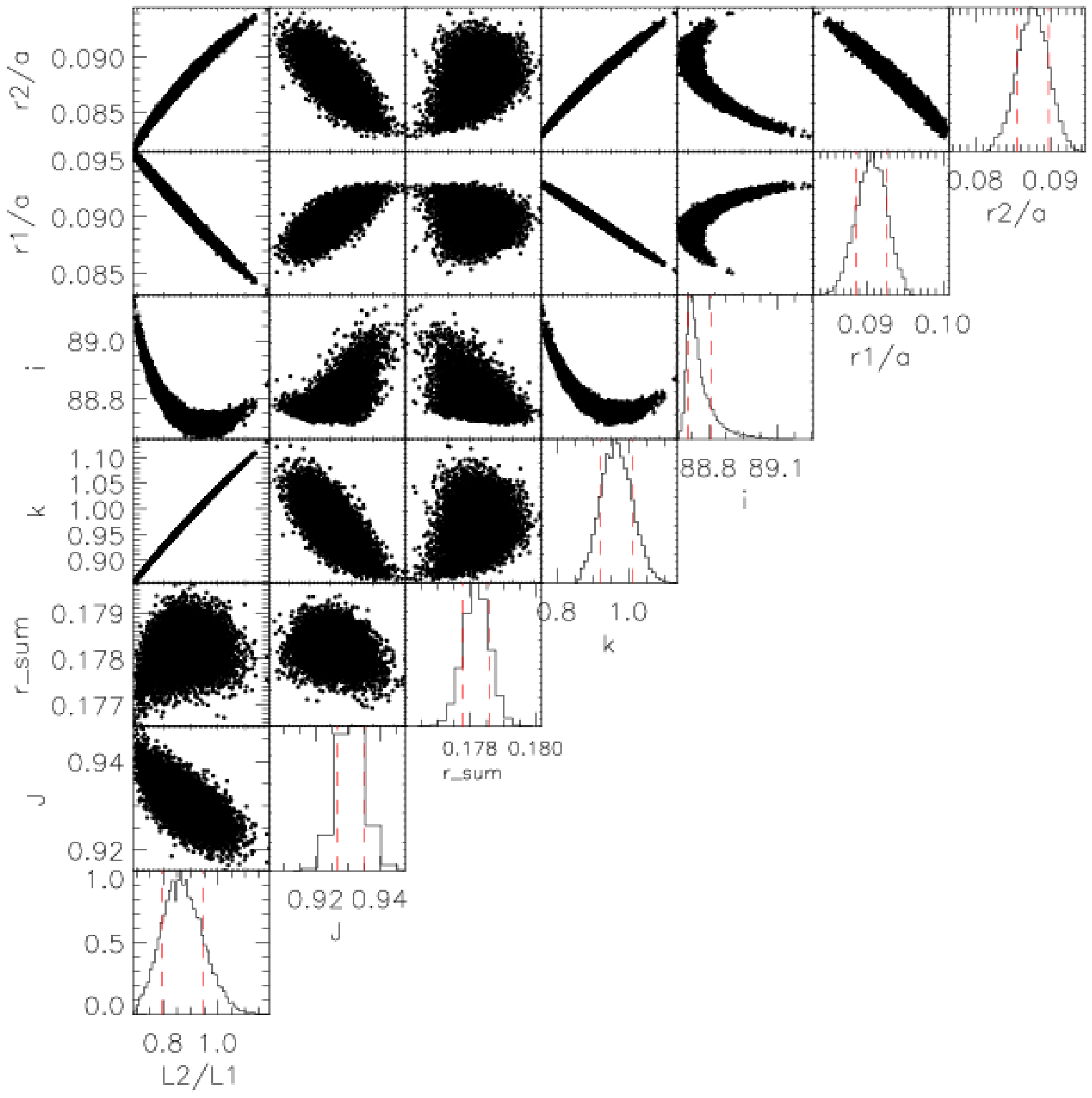}
  \caption[19b-2-01387]{{\bf 19b-2-01387} Left top panel: full phase
    WFCAM $J$-band light curve. Left bottom panel: the INT/WFC
    $i$-band light curve at secondary eclipse.  The solid red and
    purple lines show the best-fit from {\sc jktEBOP}. The blue data
    points in the smaller panels show the residuals after subtracting
    the model. Right: Parameter correlations from Monte Carlo
    simulations and histograms of individual parameter
    distributions. The red dashed vertical lines mark the $68.3\%$
    confidence interval. There is a strong correlation between the
    light ratio, the radius ratio, and the inclination (which is
    skewed), indicating the difficulty in constraining the model even
    with our high quality light curves.}
  \label{fig:19blc}
\end{figure*}

\begin{figure*}
  \centering
  \includegraphics[width=0.49\textwidth,clip=true, trim=0cm -8cm 0cm 0cm]{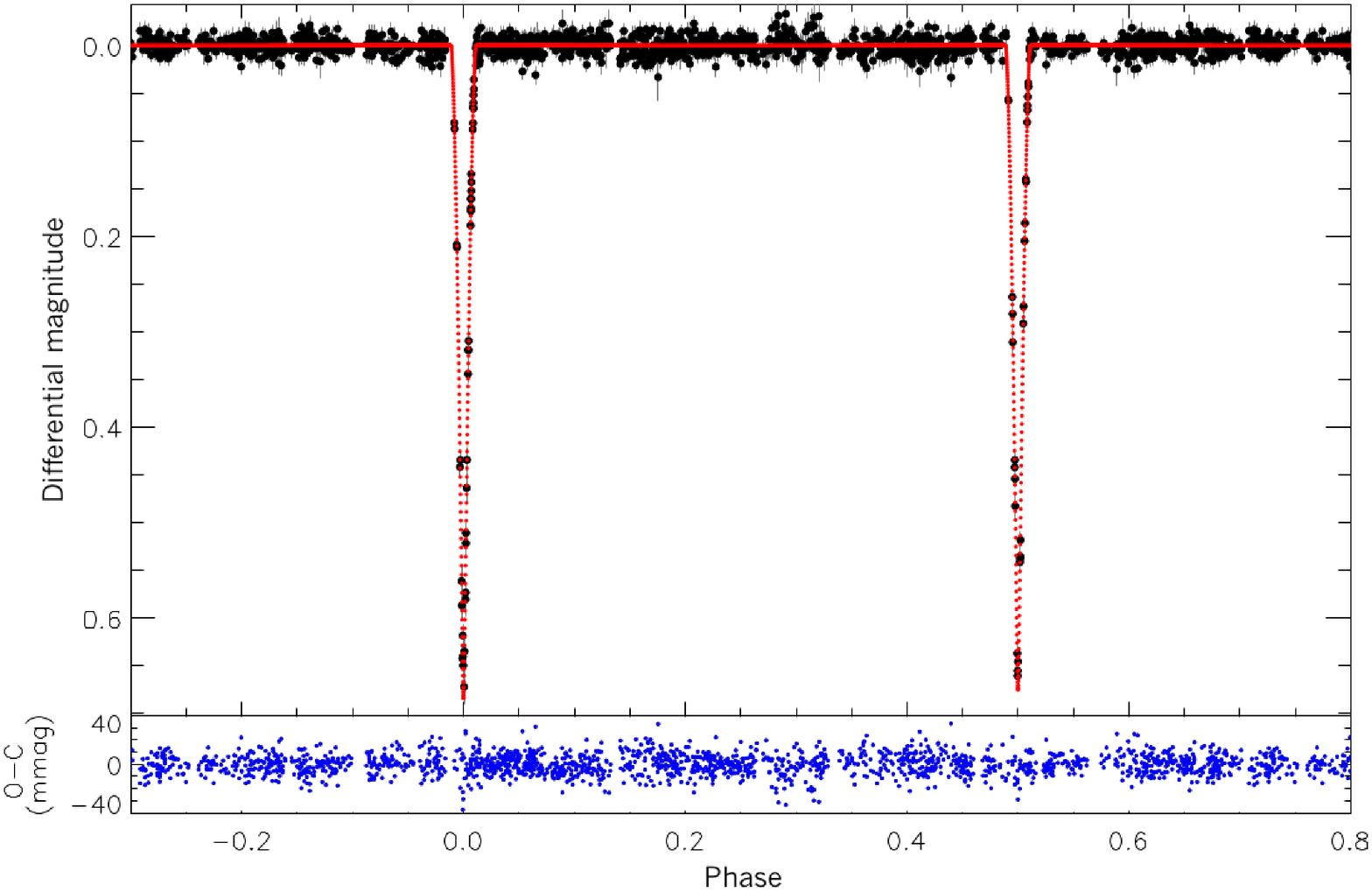}
  \includegraphics[width=0.49\textwidth]{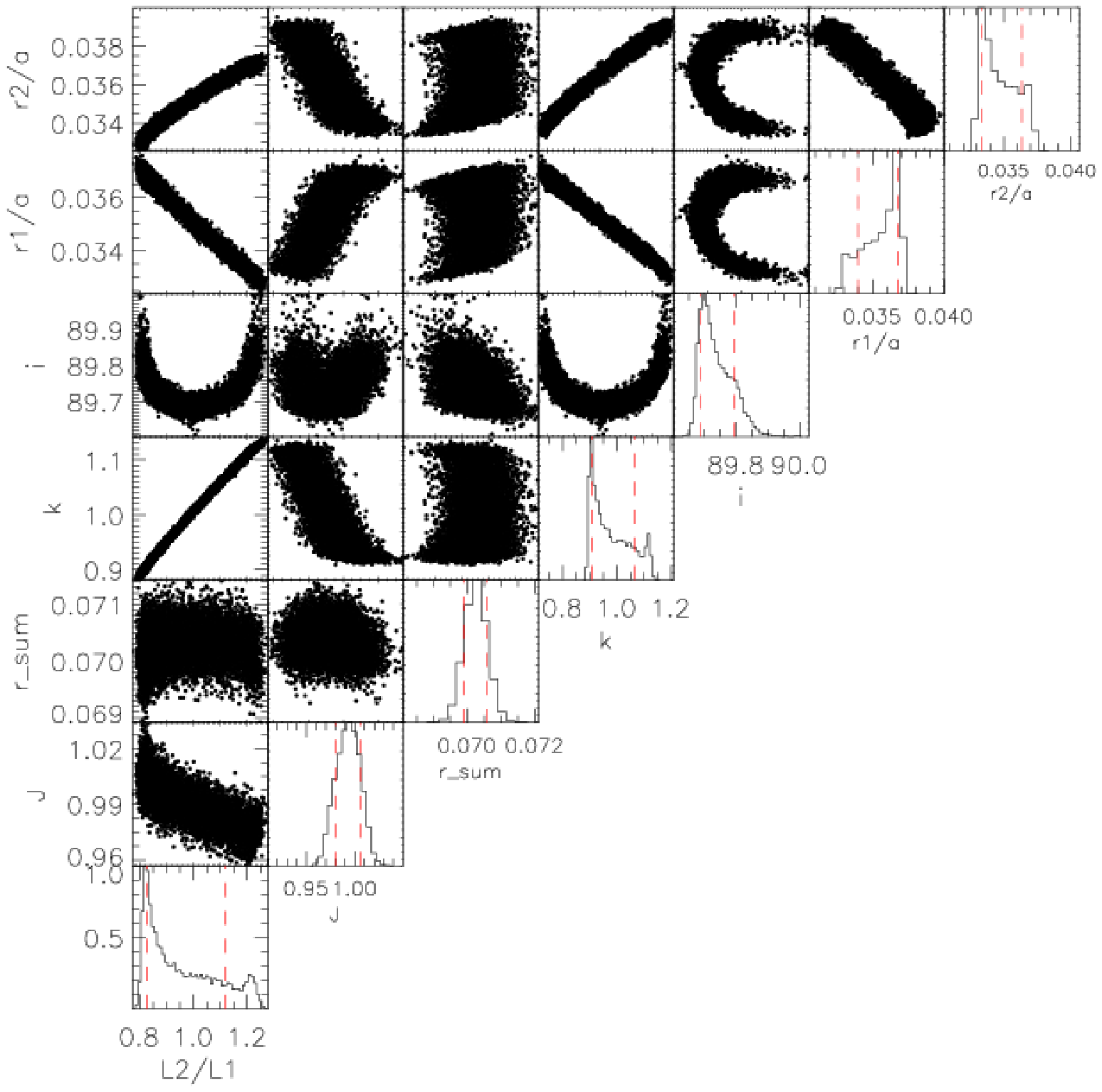}
  \caption{{\bf 19c-3-01405} Left: WFCAM $J$-band light curve. Lines
    and panels as in Figure~\ref{fig:19blc}. The magnitude scale is
    differential as we have combined light curves from two different
    WFCAM chips. Right: Monte Carlo results with lines as in
    Figure~\ref{fig:19blc}. Our inability to constrain the model with
    follow-up data results in strong correlation between the radius
    ratio and light ratio and parameter distributions that are
    significantly skewed. There are also degeneracies in the
    inclination which is expected given the near identical eclipse
    depths.}
  \label{fig:19clc}
\end{figure*}

\begin{figure*}
  \centering
    \includegraphics[width=0.49\textwidth]{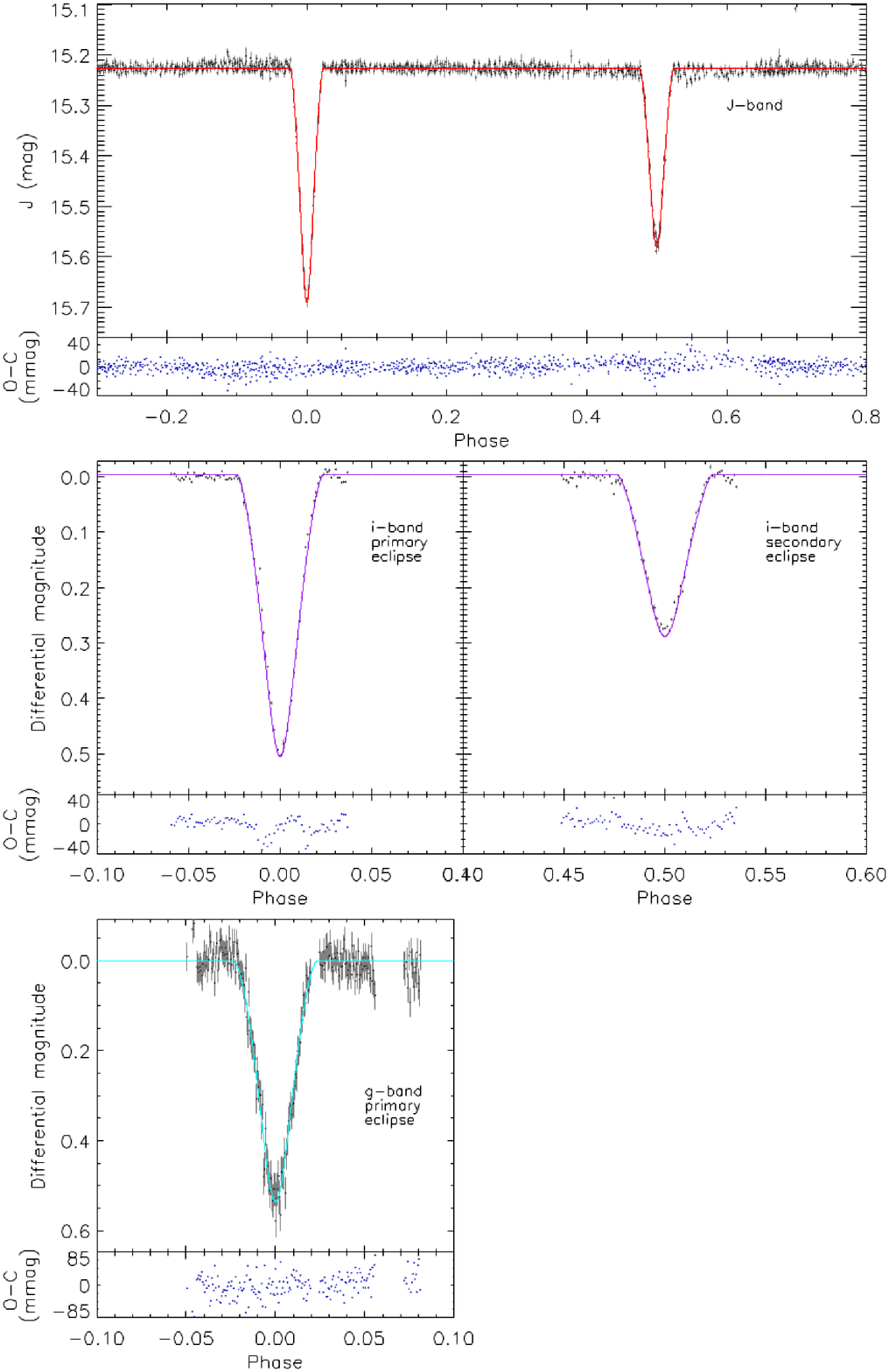}
    \includegraphics[width=0.49\textwidth,clip=true, trim=0cm -9cm 0cm 0cm]{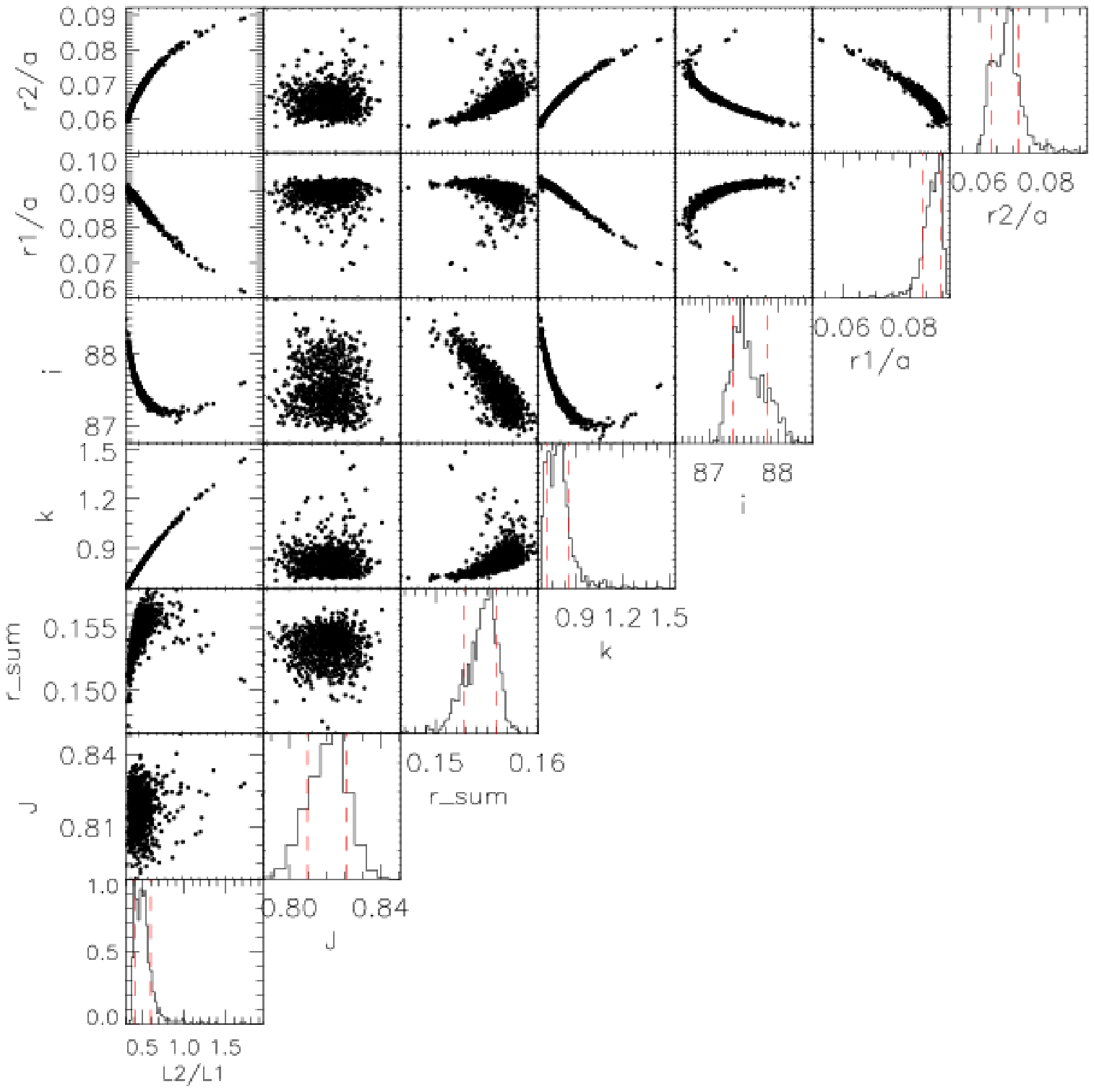}\\
    \caption{{\bf 19e-3-08413} Left top panel: full phase WFCAM
      $J$-band light curve. Left middle panel: INT/WFC $i$-band light
      curves of a primary and a secondary eclipse. Left bottom panel:
      IAC80 $g$-band light curve of a primary eclipse.  The solid red,
      purple, and cyan lines show the best-fit from {\sc
        jktEBOP}. Right: Parameter correlations from residual
      permutations, which gave the larger errors on the model
      parameters than the Monte Carlo simulations, indicating
      time-correlated systematics. There are strong correlations
      between the light ratio, radius ratio and inclination. }
  \label{fig:19elc}
\end{figure*}

The phase-folded $J$-band light curves for the MEBs and their final
model fits are shown in Figures~\ref{fig:19blc},~\ref{fig:19clc} and~\ref{fig:19elc},
while the model values are given in Table~\ref{tab:lcanal}.

\begin{table*}
\centering
\begin{tabular}{lrrrr}
\hline
\hline
Parameter&19b-2-01387&19c-3-01405&19e-3-08413\\
\hline
WTS $J$-band&&&\\
P (days)&$1.49851768\pm0.00000041$&$4.9390945\pm0.0000015$&$1.67343720\pm0.00000048$\\
$T_{0}$ (HJD)&$2454332.889802\pm0.000077$&$2454393.80791\pm0.00022$&$2454374.80821\pm0.00016$\\
($R_{1}+R_{2}$)/a&$0.17818\pm0.00040$&$0.07023\pm0.00035$&$0.1544\pm0.0016$\\
$k$&$0.967\pm0.044$&$0.987\pm0.081$&$0.782\pm0.070$\\
$J$&$0.9307\pm0.0043$&$0.993\pm0.013$&$0.8162\pm0.0084$\\
$i~(^{\circ})$&$88.761\pm0.051$&$89.741\pm0.053$&$87.59\pm0.26$\\
$e\cos\omega$&$-0.00020\pm0.00017$&$0.000060\pm0.000068$&$-0.00014\pm0.00017$\\
$e\sin\omega$&$-0.0007\pm0.0026$&$-0.0041\pm0.0059$&$0.0112\pm0.0062$\\
Normalisation (mag)&$14.64726\pm0.00017$&$0.00003\pm0.00020$&$15.22776\pm0.00020$\\
$R_{1}/a$&$0.0906\pm0.0020$&$0.0354\pm0.0014$&$0.0867\pm0.0027$\\
$R_{2}/a$&$0.0875\pm0.0021$&$0.0348\pm0.0015$&$0.0676\pm0.0040$\\
$L_{2}/L_{1}$&$0.871\pm0.076$&$0.97\pm0.15$&$0.503\pm0.090$\\
$e$&$0.0066\pm0.0026$&$0.0058\pm0.0043$&$0.0114\pm0.0062$\\
$\omega~(^{\circ})$&$268.0\pm1.7$&$180.5\pm90.9$&$91.1\pm1.2$\\
$\sigma_{J}$ (mmag)&5.2&8.4&8.7\\
\hline
INT $i$-band&&&\\
$J$&0.8100&---&0.63\\
$\sigma_{i}$ (mmag)&5.7&---&12.1\\
\hline
IAC80 $g$-band&&&\\
$J$&---&---&0.6455\\
$\sigma_{i}$ (mmag)&---&---&29.9\\
\hline
\end{tabular}
\caption{Results from the $J$ and $i$-band light curve analysis. Only
  perturbed parameters are listed. The light curve parameter
  errors are the $68.3\%$ confidence intervals while the model values
  are the means of the $68.3\%$ confidence level boundaries, such
  that the errors are symmetric. $T_{0}$ corresponds to
  the epoch of mid-primary eclipse for the first primary eclipse in
  the $J$-band light curve. Errors on 19e-3-08413 are from 
  residual permutation analysis as they were the largest, indicating
  time-correlated systematics. $\sigma_{J,i}$ give the RMS
  of the residuals to the final solutions, where all parameters in the fit are fixed to
  the quoted values and the reflection coefficients calculated from
  the system geometry.}
   \label{tab:lcanal}
\end{table*}

\subsection{Error analysis} {\sc jktEBOP} uses a Levenberg--Marquardt
minimisation algorithm \citep{Pre92} for least-squares optimisation of
the model parameters; however, the formal uncertainties from
least-squares solutions are notorious for underestimating the errors
when one or more model parameters are held fixed, due to the
artificial elimination of correlations between parameters. {\sc
  jktEBOP} provides a method for assessing the $1\sigma$ uncertainties
on the measured light curve parameters through Monte Carlo (MC)
simulations. In these simulations, a synthetic light curve is
generated using the best-fitting light curve model at the phases of
the actual observations. Random Gaussian noise is added to the model
light curve which is then fitted in the same way as the data. This
process is repeated many times and distribution of the best fits to
the synthetic light curves provide the $1\sigma$ uncertainties on each
parameter. \citet{Sou05} showed this technique is robust and gives
similar results to Markov Chain Monte Carlo techniques used by others,
under the (reasonable) assumption that the best fit to the
observations is a good fit. {\sc jktEBOP} can also perform a residual
permutation (prayer bead) bead error analysis which is useful for
assigning realistic errors in the presence of correlated noise
\citep{Sou08}. For each MEB, we have performed both MC simulations
(using $10,000$ steps) and a prayer bead analysis. The reported errors
are those from the method that gave the largest uncertainties. The
correlations between the parameter distributions from the MC and
prayer bead analysis are shown in
Figures~\ref{fig:19blc},~\ref{fig:19clc} and~\ref{fig:19elc} along
with histograms of the distributions of individual parameters. The
distributions are not perfectly Gaussian and result in asymmetric
errors for the $68.3\%$ confidence interval about the median. As we
wish to propagate these errors into the calculation of absolute
dimension, we have symmetrized the errors by adopting the mean of the
$68.3\%$ boundaries (the $15.85\%$ and $84.15\%$ confidence limits) as
the parameter value and quoting the $68.3\%$ confidence interval as
the $\pm1\sigma$ errors. These errors are given in
Table~\ref{tab:lcanal} for each MEB.

Our follow-up $g$- and $i$-band light curves (where available) were
used to check the $J$-band solution by modelling them with the derived
$J$-band parameters, but allowing the surface brightness ratio and the
light curve normalisation factor to vary. The limb darkening
coefficients were changed to those appropriate for the respective $g$-
and $i$-band and the reflection coefficients were again determined by
the system geometry. The RMS values of the these fits are given in
Table~\ref{tab:lcanal} along with the derived $g$- and $i$-band
surface brightness ratio for completeness. The $g$- and $i$-band
phase-folded data is shown overlaid with the models in
Figure~\ref{fig:19blc} and~\ref{fig:19elc}. We find that the $J$-band
solutions are in good agreement with the $g-$ and $i$-band data.

\subsection{Light ratios}
All three of our MEBs exhibit near equal-depth eclipses, implying that
the systems have components with similar mass. This is promising
because it suggest relatively large reflex motions that will appear as
well-separated peaks in a cross-correlation function from which we
derive RVs. However, it is well-known for systems with equal size
components that the ratio of the radii, which depends on the depth of
the eclipses, is very poorly determined by the light curve
\citep{Pop84}, even with the high photometric precision and large
number epochs in the WTS (see \citet{And80,Sou07d} for other excellent
examples of this phenomenon). Conversely, $(R_{1}+R_{2})/a$, is often
very well-constrained because it depends mainly on the duration of the
eclipses and the orbital inclination of the system. The reason that
the ratio of the radii is so poorly constrained stems from the fact
that quite different values of $R_{2}/R_{1}$ result in very similar
eclipse shapes.

Unfortunately, we found that all three of our MEBs presented problems
associated with poorly constrained $R_{2}/R_{1}$, revealed in the
initial modelling as either a large skew in the errors on the best-fit
parameters or best-fit solutions that were physically implausible. For
example, for 19b-2-01387, the initial best-fit gave $L_{2}/L_{1}>1$
and $R_{2}/R_{1}>1$ while $T_{2}/T_{1}<1$. We know from our
low-resolution spectroscopy that these stars are very likely to be
ordinary main-sequence M-dwarfs and while their exact radii may be
under-estimated by models, they generally obey the trend that less
massive stars are less luminous, smaller and cooler. We note that
\citet{Sta07} found a temperature reversal in a system of two young
brown dwarfs where the less massive component was hotter but smaller
and fainter than its companion. In their case the more massive
component, although cooler, had an RV curve and eclipse depth that
were consistent. In our cases, the most massive component (smallest
$K_{\star}$) comes towards us (blue-shift) after the deepest (primary)
eclipse, so it must be the more luminous component. The uncertainty in
our modelling is most likely to due to insufficient coverage of the
mid-eclipse points.

However, we can try to use external data as an additional constraint
in the fit. Some authors employ a spectroscopically derived light
ratio as an independent constraint on $k$ in the light curve modelling
\citep{Sou04a,Sou07,Nor94}. {\sc jktEBOP} allows the user to
incorporate an input light ratio in the model and propagates the
errors in a robust way. The input light ratio adds a point in the flux
array at a specific phase \citep{Sou07}. If this is supplied with a
very small error, the point is essentially fixed. We have tried
several methods to estimate the light ratio for each of our three
systems, although we stress here that none of the estimates should be
considered as significant. One requires high resolution spectra to
extract precise light ratios, via the analysis of the equivalent width
ratios of metallic lines, which will be well-separated if observed at
quadrature \citep{Sou05}. With a high resolution spectrum, one can
disentangle the components of the eclipsing binary and perform
spectral index analysis on the separate components (e.g.
\citealt{Irw07b}).

\begin{figure}
  \centering
  \includegraphics[width=0.5\textwidth]{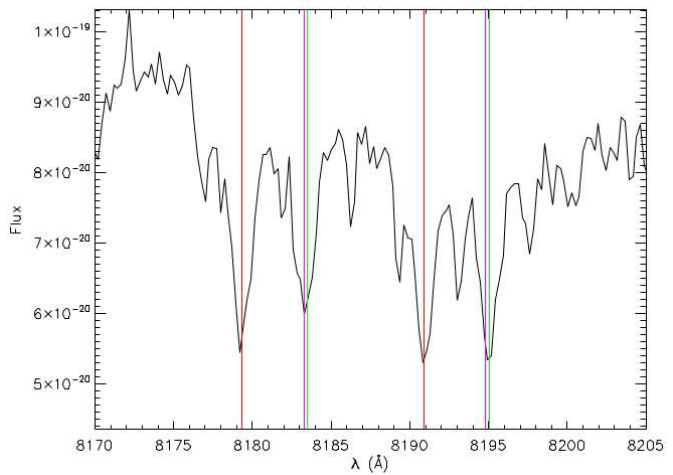}
  \caption{{\bf 19b-2-01387}: A high signal-to-noise intermediate
    resolution spectrum taken near quadrature phase of 19b-2-01387 in
    the Na II doublet wavelength region which we used to measure
    equivalents widths to estimate the light ratio. The purple
    vertical lines show the rest frame wavelength of the doublet at
    $\lambda8183.27$\AA, $\lambda8194.81$\AA. The red lines mark
    the doublet for primary object and the green lines mark the
    secondary doublet lines, based on the RVs derived in
    Section~\ref{sec:RVs}.}
  \label{fig:Na_doublet}
\end{figure}

19b-2-01387 is our brightest system and subsequently has the highest
signal-to-noise in our intermediate-resolution spectra. The best
spectrum is from the first night of observations. For this system, we
estimated the light ratio in three ways: i) by measuring the ratio of
the equivalent widths of the lines in the Na II doublet (shown in
Figure~\ref{fig:Na_doublet}), ii) by using the two-dimensional
cross-correlation algorithm, {\sc todcor} \citep{Zuc94}, which weights
the best-matching templates by the light ratio and, iii) by
investigating the variation in the goodness-of-fit for a range of
input light ratios in the model.

For the first method, the {\sc iraf.splot} task was used to measure
the equivalent width of the Na I doublet feature with rest wavelength
$8183.27$\AA ~for each star. Note that this assumes the components
have the same effective temperature. The ratio was
$EW(2)/EW(1)=0.3582/0.4962 = 0.7219$. In the second method, we found
that only the spectrum from the first night contained sufficient SNR
to enable {\sc todcor} to correctly identify the primary and secondary
components. It is known that {\sc todcor} does not perform as well for
systems with similar spectral features \citep{Sou07a} so we do not use
it to derive RVs for our nearly equal mass systems. The {\sc todcor}
estimated light ratio was $L_{2}/L_{1}=0.846$. In the final method, we
iterated {\sc jktEBOP} across a grid of initial light ratios between
0.6-1.1, in steps of 0.01, with very small errors while allowing all
our usual parameters to vary. The resulting $\chi^{2}$-distribution is
not well-behaved. There is a local and global minimum at
$L_{2}/L_{1}=0.72$ and $L_{2}/L_{1}=0.97$, respectively, but the
global minimum is bracketed on one side by a significant jump to a
much larger $\chi^{2}$ suggesting numerical issues. We opted to use
the light ratio derived with {\sc
  todcor} as the input to the model. This value lies half-way between
the two minimums of the $\chi^{2}$ distribution, so we supplied it
with a $\sim15\%$ error to allow the parameter space to be explored,
given the uncertainty in our the measurement. Our follow-up $i$-band
data of a single secondary eclipse also prefers a light ratio less
than unity, but the lack of phase coverage does not give a
well-constrained model. The resulting parameter distributions, shown
in Figure~\ref{fig:19blc}, show strong correlation between the light
ratio and $R_{2}/R_{1}$ as expected. The resulting $1\sigma$ error
boundary for the light ratio, which is computed from $k$ and $J$, is
in broad agreement with the methods used to estimate it.

For 19e-3-08413, we obtained additional $i$-band photometry of a
primary and secondary eclipse, plus a further primary eclipse in the
$g$-band. Here, we have estimated the light ratio by fitting our two
datasets in these wavebands separately, using appropriate limb
darkening coefficients for the $i$- and $g$-bands in each case, and
allowing all our usual parameters to vary. We find best-fit values
from the $i$- and $g-$bands of $L_{2}/L_{1}=0.29$ and
$L_{2}/L_{1}=0.36$, respectively. This confirms a light ratio less
than unity, but as the light ratio depends on the surface brightness
ratio, which in turn is wavelength dependent, we adopted
$L_{2}/L_{1}=0.29$ with input with an error of $\pm0.5$ in the final
fit to the $J$-band data. Note we chose to use the $i$-band value as
it is closer in wavelength to the $J$-band and the light curve was of
higher quality.

In the case of 19c-3-01405, we could not derive a light ratio from the
low SNR spectra, nor do we have follow-up $i$-band photometry (due to
time scheduling constraints). The eclipses are virtually identical so
we supplied an input light ratio of $L_{2}/L_{1}=1.0$ with an error of
$50\%$. Unfortunately, the final error distributions for the
parameters are still quite skewed, as shown in Figure~\ref{fig:19clc}.

\subsection{Star spots}
For 19e-3-08413, we found that the residual permutation analysis gave
larger errors, indicating time-correlated systematics. We have not
allowed for spot modulation in our light curve model thus the
residuals systematics may have a stellar origin. As mentioned
previously, we expect star spot modulation in the $J$-band to be
relatively weak as the SED of the spot and the star at these
wavelength are more similar than at shorter wavelengths. It is
difficult to test for the presence of spots in the $g$- and $i$-band
data as we do not have suitable coverage out-of-eclipse. We only have
full-phase out-of-eclipse observations in a single $J$-bandpass
therefore any physical spot model will be too degenerate between
temperature and size to be useful. Furthermore, our $J$-band data span
nearly four years, yet spot size and location are expected to change
on much shorter timescales, which leads to a change in the amplitude
and phase of their sinusoidal signatures. Stable star spot signatures
over the full duration of our observations would be unusual. The WTS
observing pattern therefore makes it difficult to robustly fit simple
sinusoids, as one would need to break the light curve into many
intervals in order to have time spans where the spots did not change
significantly (e.g. three week intervals), and these would
consequently consist of few data points. Nevertheless, we have
attempted to test for spot modulation in a very simplistic manner by
fitting the residuals of our light curve solutions as a function of
time ($t$) with the following sinusoid:

\begin{equation}
f(t)=a_{0}+a_{1}\sin(2\pi (t/a_{2}) + a_{3}),
\label{eqn:sine}
\end{equation}

where the systemic level ($a_{0}$), amplitude ($a_{1}$), and phase
($a_{3}$) were allowed to vary in the search for the best-fit, while
the period ($a_{2}$) was held fixed at the orbital period as we expect
these systems to be synchronised (see Table~\ref{tab:dimensions} for
the theoretical synchronisation timescales). Once the best-fit was
found, the values were used as starting parameters for the {\sc idl}
routine {\sc mpfitfun}, to refine the fit and calculate the errors on
each parameter. Table~\ref{tab:sine} summarises our findings.

\begin{table*}
\centering
\begin{tabular}{lrrrrrrrr}
\hline
\hline
Name&Amplitude&Phase&$\gamma$&$\chi^{2}_{\nu,\rm  before}$&$\chi^{2}_{\nu,\rm
  after}$&RMS$_{\rm before}$&RMS$_{\rm after}$\\
&(mmag)&&(mmag)&&&(mmag)&(mmag)\\
\hline
19b-2-01387&$1.83\pm0.23$&$2.53\pm0.012$&$0.19\pm0.15$&1.11&1.04&5.2&4.9\\
19c-3-01405&$0.22\pm0.27$&$-1.5\pm1.3$&$0.23\pm0.20$&0.87&0.87&8.4&8.4\\
19e-3-08413&$3.47\pm0.32$&$-0.143\pm0.050$&$0.39\pm0.22$&1.32&1.19&7.8&7.5\\
\hline
\end{tabular}
\caption{Results of modelling the light curve model residuals with the
  simple sinusoid defined by Equation~\ref{eqn:sine}, to test for the
  presence of spot modulation. The terms `before' and `after' refer to the
  reduced $\chi^{2}$ and RMS values before subtracting the best-fit sine curve and then after the  subtraction. Note: mmag = $10^{-3}$ mag. The RMS$_{\rm before}$
  value for 19e-3-08413 is different to Table~\ref{tab:lcanal} as one
  data point was clipped due to it being a significant outlier.}
\label{tab:sine}
\end{table*}

There is evidence to suggest a low-level synchronous sinusoidal
modulation in 19b-2-01387 and 19e-3-08413 with amplitude $\sim1.8-3.5$
mmag, but we do not find significant modulation for our longest period
MEB (19c-3-01405). The modulation represents a source of systematic
error that if modelled and accounted for, could reduce the errors our
radius measurements. However, with only one passband containing
out-of-eclipse variation, we cannot provide a useful non-degenerate
model. Good-quality out of eclipse monitoring is required and given
that spot modulation evolves, contemporaneous observations are needed,
preferably taken at multiple wavelengths to constrain the spot
temperatures \citep{Irw11}. It is surprising that the apparent spot
modulation in our MEBs persists over the long baseline of the WTS
observations and perhaps an alternate explanation lies in residual
ellipsoid variations from tidal effects between the two stars. We note
here that our limiting errors in comparing these MEBs to the
mass-radius relationship in Section~\ref{sec:mrrel} are on the masses,
not the radii.

\section{Radial velocity analysis}\label{sec:RVs}
To extract the semi-amplitudes ($K_{1}, K_{2}$) and the centre-of-mass
(systemic) velocity, $\gamma$, of each MEB system, we modelled the RV
data using the {\sc idl} routine {\sc mpfitfun} \citep{Mark09}, which uses the
Levenberg--Marquardt technique to solve the least-squares problem. The
epochs and periods were fixed to the photometric solution values as
these are extremely well-determined from the light curve. Circular
orbits were assumed ($e=0$) for all three systems as the eccentricity
was negligible in all light curve solutions. We fitted the primary RV
data first using the following model:

\begin{equation}
RV_{1} = \gamma - K_{1}\sin(2\pi\phi)
\end{equation}

where $\phi$ is the phase, calculated from the light curve solution,
and $K$ is the semi-amplitude. To obtain $K_{2}$, we then fitted the
secondary RV data points using the equation above, but this time fixed
$\gamma$ to the value determined from the primary RV data.

\begin{equation}
RV_{2} = \gamma + K_{2}sin(2\pi\phi)
\end{equation}

The errors on each RV measurement are weighted by the RV error given
by {\sc iraf.fxcor} and then scaled until the reduced $\chi^{2}$ of
the model fit is unity. The RMS of the residuals is quoted alongside
the derived parameters in Table~\ref{tab:rv_anal}, and is treated as
the typical error on each RV data point. The RMS ranges from $\sim2-5$
km/s between the systems and for the given magnitudes of our systems
is the same as the predictions of \citet{Aig07} who calculated the
limiting RV accuracy for ISIS on the WHT, when using 1 hour exposures
and an intermediate resolution grating centred on $8500$\AA.

\begin{figure}
  \centering
  \includegraphics[width=0.49\textwidth]{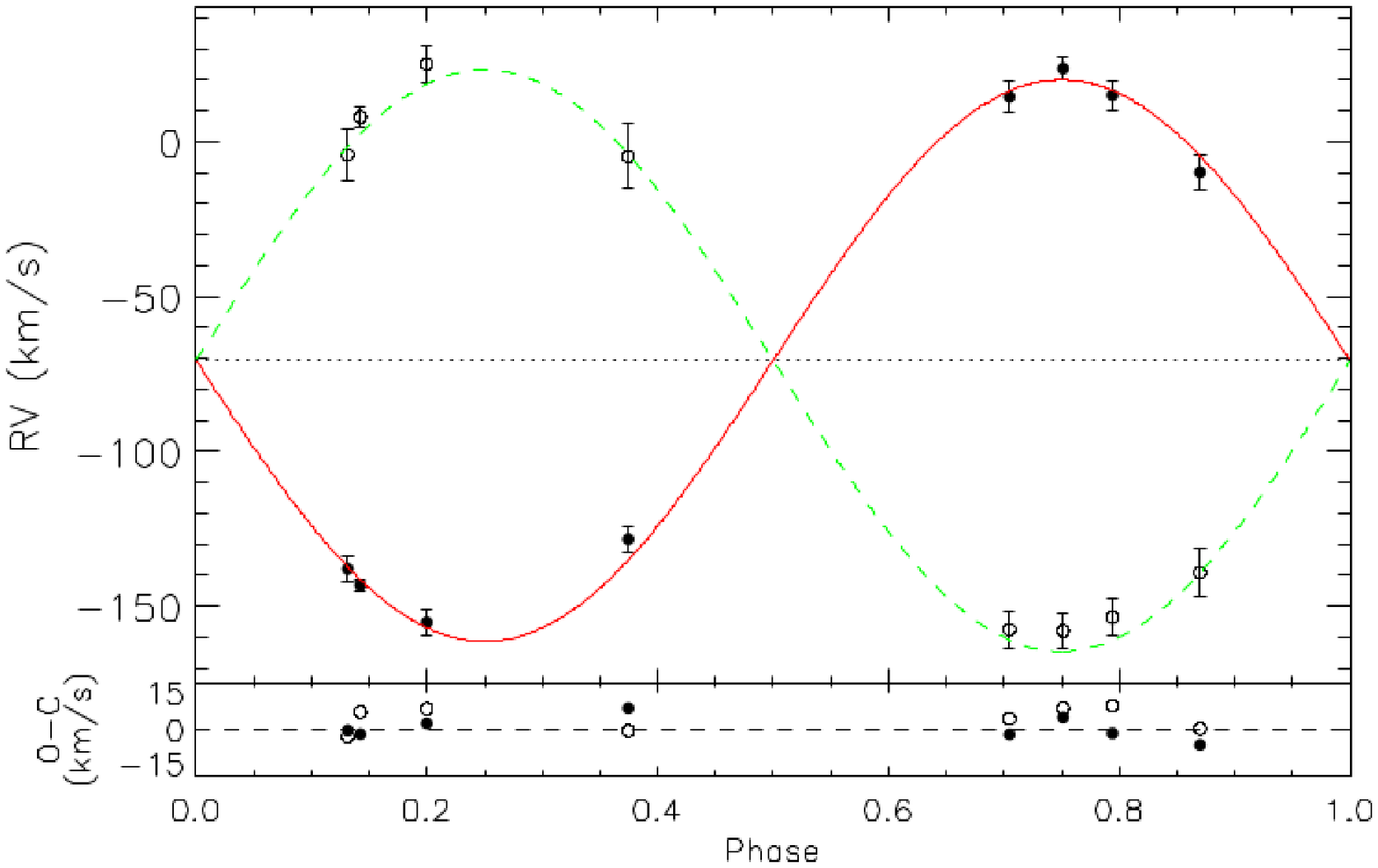}\\
  \includegraphics[width=0.49\textwidth]{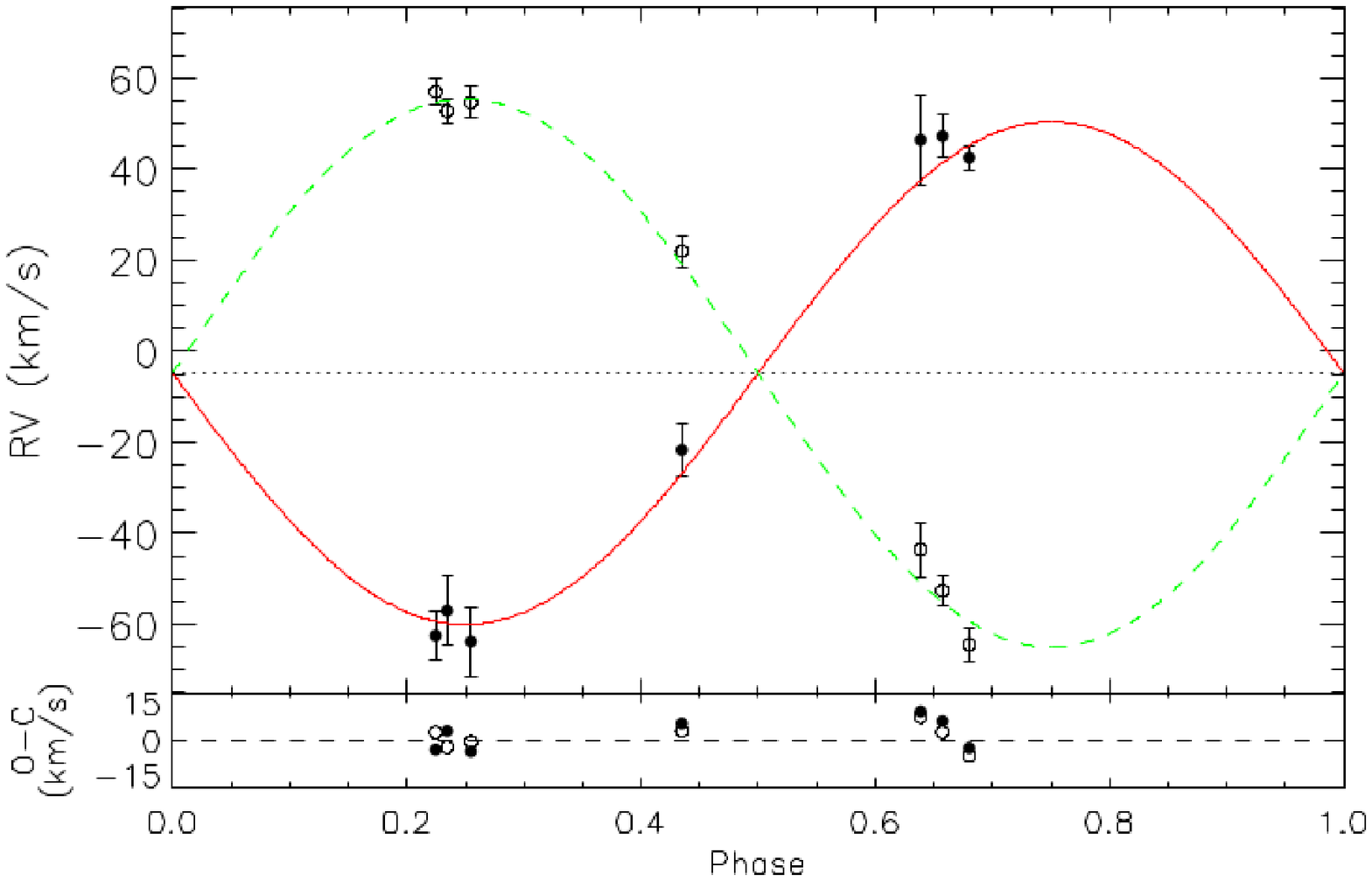}\\
  \includegraphics[width=0.49\textwidth]{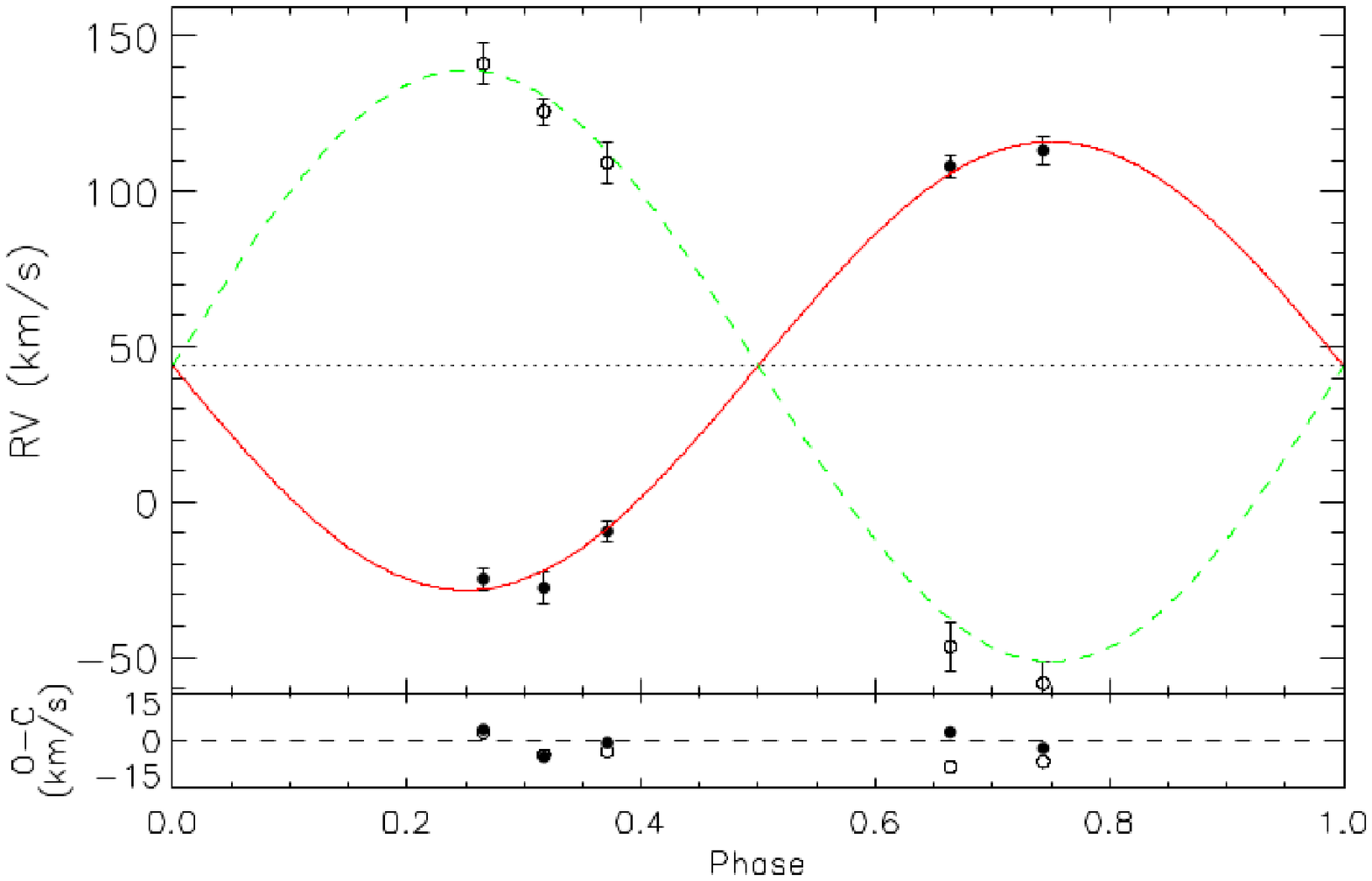}
  \caption{Primary and secondary RV curves for the MEBs. Top: {\bf
      19b-2-01387}; Middle: {\bf 19c-3-01405}; Bottom: {\bf
      19e-3-08413}. The solid black circles are RV measurements for
    the primary star, while open circles denote the secondary star RV
    measurements. The solid red lines are the model fits to the
    primary RVs and the dashed green lines are the fits to the
    secondary RVs, fixed to the systemic velocity of their respective
    primaries. The horizontal dotted lines mark the systemic
    velocities. The error bars are from {\sc iraf.fxcor} but are
    scaled so that the reduced $\chi^{2}$ of the model fit is unity.
    They are merely an indication of the signal-to-noise of the individual
    spectra and the mismatch between the template and data.
    Under each RV plot is a panel showing the residuals of
    the best-fits to the primary and secondary RVs. Note the change in
    scale for the y-axis. The typical RV error for each component is
    given in Table~\ref{tab:rv_anal} by the RMS of their respective
    residuals.}
  \label{fig:RVs}
\end{figure}

The RV curves for the primary and secondary components
of the three MEBs are shown in Figure~\ref{fig:RVs} along with the
residuals of each fit. The error bars are the scaled errors from {\sc
  iraf.fxcor} and serve as an indicator of the signal-to-noise in the
individual spectra and the degree of mismatch with the best template.

\begin{table}
\centering
\begin{tabular}{lrrr}
  \hline
  \hline
  Parameter&19b-2-01387&19c-3-01405&19e-3-08413\\
  \hline
  $K_{1}$ (km/s)&$90.7\pm1.6$&$55.2\pm2.2$&$72.1\pm2.0$\\
  $K_{2}$ (km/s)&$94.0\pm2.3$&$60.2\pm1.4$&$95.2\pm3.0$\\
  $\gamma$ (km/s)&$-70.7\pm1.3$&$-4.8\pm2.0$&$43.8\pm1.8$\\
  $\rm RMS_{1}$ (km/s)&1.8&3.7&2.7\\
  $\rm RMS_{2}$ (km/s)&5.4&2.5&5.0\\
  $q$                                       &$0.965\pm0.029$&$0.917\pm0.042$&$0.757\pm0.032$\\
  $a\sin i$ $(\rsun)$             &$5.472\pm0.083$&$11.27\pm0.25$  &$5.53\pm0.12$\\
  $M_{1}\sin^{3}i$ ($\msun$)&$0.498\pm0.019$&$0.410\pm0.021$&$0.462\pm0.025$\\
  $M_{2}\sin^{3}i$ ($\msun$)&$0.480\pm0.017$&$0.376\pm0.023$&$0.350\pm0.018$\\
\hline
\end{tabular}
   \caption[RV Analysis]{Results from radial velocity analysis.}
   \label{tab:rv_anal}
\end{table}

\section{Absolute dimensions and space velocities}\label{sec:absdim}
Combining the results of the light curve and RV curve modelling allows
us to derive the absolute masses and radii of our MEB components.
Table~\ref{tab:dimensions} gives these dimensions along with the
separations, individual effective temperatures, surface gravities, and
bolometric luminosities for each binary system. The masses and radii
lie within the ranges $0.35-0.50\msun$ and $0.37-0.5\rsun$
respectively, and span orbital periods from $1-5$ days. The derived
errors on the masses and radii are $\sim3.5-6.4\%$ and
$\sim2.7-5.5\%$, respectively.

\begin{table}
\centering
\begin{tabular}{@{\extracolsep{\fill}}l@{\hspace{2pt}}r@{\hspace{7pt}}r@{\hspace{7pt}}r@{\hspace{1pt}}}
  \hline
  \hline
  Parameter&19b-2-01387&19c-3-01405&19e-3-08413\\
  \hline
  $M_{1}$ ($\msun$)              &$0.498\pm0.019$&$0.410\pm0.023$&$0.463\pm0.025$\\
  $M_{2}$ ($\msun$)              &$0.481\pm0.017$&$0.376\pm0.024$&$0.351\pm0.019$\\
  $R_{1}$ ($\rsun$)                &$0.496\pm0.013$&$0.398\pm0.019$&$0.480\pm0.022$\\
  $R_{2}$ ($\rsun$)                &$0.479\pm0.013$&$0.393\pm0.019$&$0.375\pm0.020$\\
  a ($\rsun$)                          &$5.474\pm0.083$&$11.27\pm0.27$&$5.54\pm0.12$\\
  $\log(g_{1})$&$4.745\pm0.039$ &$4.851\pm0.055$&$4.742\pm0.053$\\
  $\log(g_{2})$&$4.760\pm0.035$ &$4.825\pm0.064$&$4.834\pm0.051$\\
  $\rm T_{\rm eff,1}$ (K)         &$3498\pm100$     &$3309\pm130$    &$3506\pm140$\\
  $\rm T_{\rm eff,2}$ (K)         &$3436\pm100$     &$3305\pm130$    &$3338\pm140$\\
$L_{\rm bol,1} (\rm L_{\sun})$&$0.0332\pm0.0042$&$0.0172\pm0.0031$&$0.0314\pm0.0058$\\
$L_{\rm bol,2} (\rm L_{\sun})$&$0.0289\pm0.0037$&$0.0166\pm0.0031$&$0.0167\pm0.0033$\\
$M_{\rm 1,bol}$                      &$8.45\pm0.14$&$9.16\pm0.20$&$8.51\pm0.19$\\
$M_{\rm 2,bol}$                      &$8.60\pm0.14$&$9.20\pm0.20$&$9.26\pm0.23$\\
$V_{\rm 1rot,sync}$ (km/s)     &$16.73\pm0.45$&$4.08\pm0.19$&$14.51\pm0.55$\\
$V_{\rm 2rot,sync}$ (km/s)     &$16.73\pm0.45$&$4.01\pm0.20$&$11.31\pm0.70$\\
$t_{\rm sync}$ (Myrs)             &$0.05$&$6.3$&$0.1$\\
$t_{\rm circ}$ (Myrs)              &$2.6$&$1480$&$4.0$\\
$d_{\rm adopted}$ (pc)           &$545\pm29$&$645\pm53$&$610\pm52$\\
U (km/s)                               &$-63.6\pm7.0$&$-2.4\pm9.0$&$30.9\pm8.6$\\
V (km/s)                                &$1.0\pm7.8$&$1.3\pm12.2$&$-10.2\pm11.8$\\
W (km/s)                               &$-37\pm6.4$&$-4.2\pm8.5$&$30.1\pm8.1$\\
  \hline
\end{tabular}
\caption{Derived properties for the three MEBs. $V_{\rm rot,sync}$ are
  the rotational velocities assuming the rotation period is synchronised
  with the orbital period. $t_{\rm sync}$ and $t_{\rm circ}$ are the
  theoretical tidal synchronisation and circularisation timescales
  from \citet{Zah75,Zah77}}
   \label{tab:dimensions}
\end{table}

Eclipsing binaries are one of the first rungs on the Cosmic Distance
Ladder and have provided independent distance measurements within the
local group e.g. to the Large Magellanic Cloud and to the Andromeda
Galaxy \citep{Gui98,Rib05,Bon07}. The traditional method for measuring
distances to eclipsing binaries is to compute the bolometric magnitude
using the luminosity, radius and effective temperature found from the
light curve and RV curve analysis. This is combined with a bolometric
correction and the system apparent magnitude to compute the
distance. While this can yield quite accurate results, the definitions
for effective temperature and the zero points for the absolute
bolometric magnitude and the bolometric correction must be consistent
\citep{Bes98,Gir02}. However, we have opted to use a different method
to bypass the uncertainties attached to bolometric corrections. We
used {\sc jktabsdim} \citep{Sou05b}, a routine that calculates
distances using empirical relations between surface brightness and
effective temperature. These relations are robustly tested for dwarfs
with $\teff>3600$ K and there is evidence that they are valid in the
infrared to $\sim3000$ K \citep{Ker04}. The scatter around the
calibration of the relations in the infrared is on the $1\%$
level. The effective temperature scales used for the EB analysis and
the calibration of its relation with surface brightness should be the
same to avoid systematic errors but this is a more relaxed constraint
than required by bolometric correction methods \citep{Sou05b}. The
infrared $J, H$ and $K$-bands are relatively unaffected by
interstellar reddening but we have shown in Section~\ref{sec:redden}
that we expect a small amount. In the distance determination, we have
calculated the distances at zero reddening and at the maximum
reddening ($E(B-V)=0.103$ at 1 kpc for early M-dwarfs with
$J\leq16$ mag). Our adopted distance, $d_{\rm adopted}$, reported in
Table~\ref{tab:dimensions} is the mid-point of the minimum and maximum
distance values at the boundaries of their the individual errors,
which includes the propagation of the effective temperature
uncertainties. The MEBs lie between $\sim550-650$ pc.

With a full arsenal of kinematic information (distance, systemic
velocities, proper motions and positions) we can now derive the true
space motions, $UVW$, for the MEBs and determine whether they belong
to the Galactic disk or halo stellar populations.  We used the method
of \citet{Joh87} to determine $UVW$ values with
respect to the Sun (heliocentric) but we adopt a left-handed
coordinate system to be consistent with the literature, that is, $U$
is positive away from the Galactic centre, $V$ is positive in the
direction of Galactic rotation and $W$ is positive in the direction of
the north Galactic pole. We use the prescription of \citet{Joh87} to
propagate the errors from the observed quantities and the results are
summarised in Table~\ref{tab:dimensions}.

\begin{figure}
  \centering
  \includegraphics[width=0.5\textwidth]{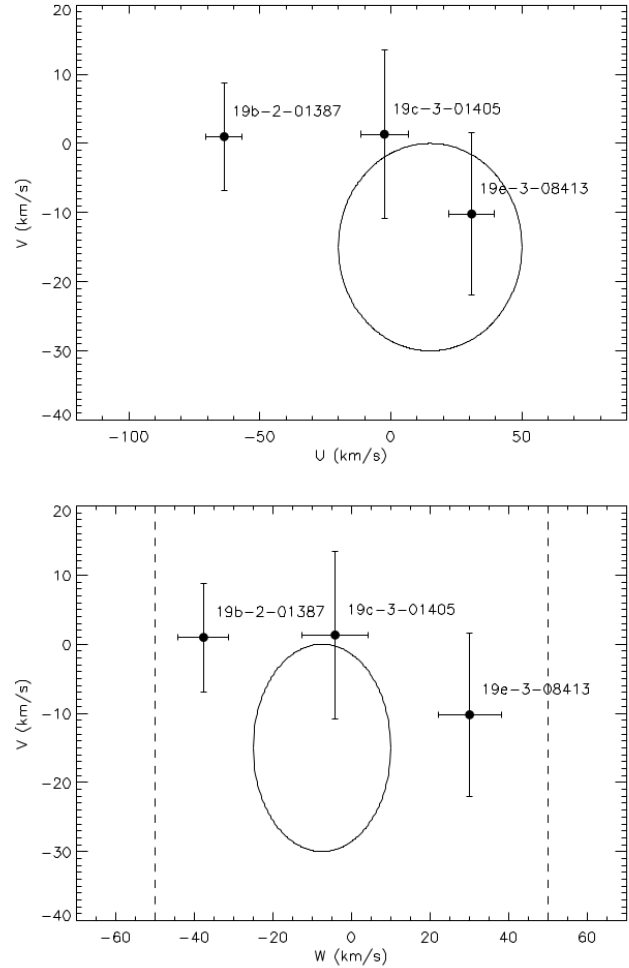}
  \caption{The UVW space motions with respect to the Sun for our
    MEBs. The errors have been propagated according to
    \citet{Joh87}. The solid ellipses are the error ellipses for the
    young disk defined by \citet{Leg92}. The dashed vertical lines in
    the lower plot mark the $W$ boundary within which the young-old
    disk population is contained \citep{Leg92}.}
  \label{fig:uvw}
\end{figure}

Figure~\ref{fig:uvw} shows the MEBs in relation to the error ellipse
for the Galactic young disk as defined by \citet{Leg92} ($-20<U<50$,
$-30<V<0$, $-25<W<10$ w.r.t the Sun). 19c-3-01405 is consistent within
its error with the young disk. 19b-2-01387 is an outlier to the young
disk criterion. Instead, \citet{Leg92} define objects around the edges
of the young disk ellipse as members of the young-old disk population,
which has a sub-solar metallicity ($-0.5<[m/H]<0.0$). 19e-3-08413
exceeds the allowed $W$ range for the young disk, despite overlap in
the $UV$ plane. \citet{Leg92} assign these objects also to the
young-old disk group. This suggest that two of our MEBs could be
metal-poor but our spectral index measurements in
Section~\ref{sec:indices} are not accurate enough to confirm this. We
would require, for example, higher resolution, $J$-band spectra to
assess the metallicities in detail \citep{One11}. Comparisons with
space motions of solar neighbourhood moving groups do not reveal any
obvious associations \citep{Sod93b}.

\section{Discussion}\label{sec:discuss}

\subsection{The mass-radius diagram}\label{sec:mrrel}
\begin{figure*}
  \centering
  \includegraphics[width=\textwidth]{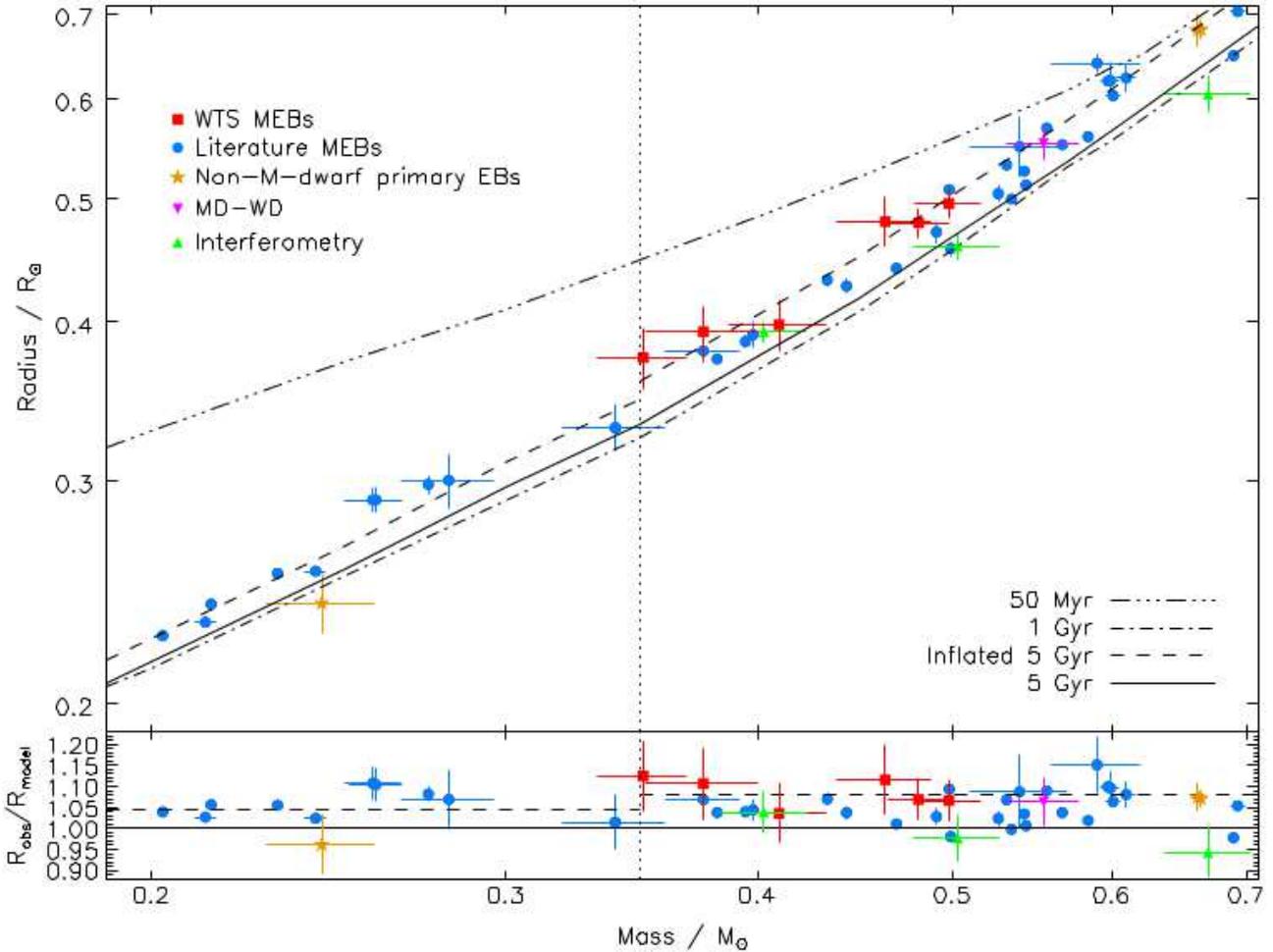}
  \caption{The mass-radius diagram for low-mass stars. The filled
    circles show literature MEB values with reported mass errors
    $<6\%$ and radius errors $<6.5\%$. Also shown are literature
    values for i) the low-mass secondaries of eclipsing binaries with
    primary masses $>0.6\msun$, ii) M-dwarfs found in M-dwarf - white
    dwarf eclipsing binaries (MD-WD), and iii) radius measurements of
    single M-dwarfs from interferometric data. The red squares mark
    the new WTS MEBs.  The diagonal lines show model isochrones from
    the \citet{Bar98} models ($[m/H]=0$, $Y=0.275$ and
    $L_{mix}=H_{P}$), while the vertical dotted line marks the onset
    of fully-convective envelopes \citep{Chab97}. The dashed line
    shows the 5 Gyr isochrone plus the average radius excess found by
    \citet{Kni11}, assuming a discontinuity at the
    fully-convective transition. Above $0.35\msun$, the model is
    inflated by $7.9\%$, but below it is only inflated by
    $4.5\%$. The bottom panel shows the radius anomaly, $R_{\rm
      obs}/R_{\rm model}$ computed using the $5$ Gyr isochrone and
    again the dashed line shows the corresponding average radius
    excess found by{Kni11}. The literature data used in these plots
    are given in Table~\ref{tab:mrptdata}.}
  \label{fig:mr-rel}
\end{figure*}

Figure~\ref{fig:mr-rel} shows the positions of our MEBs in the
mass-radius plane and compares them to literature mass-radius
measurements derived from EBs with two M-dwarfs, EBs with an M-dwarf
secondary but hotter primary, eclipsing M-dwarf - white dwarf systems,
and inactive single stars measured by interferometry. We only show
values with reported mass and radius errors comparable to or better
than our own errors. The solid line marks the $5$ Gyr, solar
metallicity isochrone from the \citet{Bar98} models (solid line), with
a convective mixing length equal to the scale height ($\rm L_{\rm
  mix}=H_{\rm P}$), while the dash-dot line shows the corresponding $1$ Gyr
isochrone.

It is clear that some MEBs, both in the WTS and in the literature,
have an excess in radius above the model predictions, and although
there is no evidence to say that all MEBs disagree with the models,
the scatter in radius at a given mass is clear, indicating a residual
dependency on other parameters. \citet{Kni11} measured the average
fractional radius excess below $0.7\msun$, but based on the findings
of \citet{Chab07} and \citet{Mor10}, split the sample at the
fully-convective boundary to investigate the effect of inhibited
convection. The dashed line in Figure~\ref{fig:mr-rel} marks the
average radius inflation they found with respect to the 5 Gyr
isochrone for the fully-convective mass region below $0.35\msun$ and
in the partially-convective region above ($7.9\%$ for $>0.35\msun$,
but only by $4.5\%$ for $>0.35\msun$). The WTS MEBs sit systematically
above the 5 Gyr isochrone but appear to have good agreement with the
average radius inflation for their mass range. It is interesting to
note that we find similar radius excesses to the literature despite
using infrared light curves. At these wavelengths, we crudely expect
lower contamination of the light curves by sinusoidal star spots
signals and less loss of circular symmetry, on account of the smaller
difference between the spectral energy distributions of the star and
the spots in the $J$-band. If one could eliminate the $\sim3\%$
systematic errors in MEB radii caused by polar star spots
\citep{Mor10} by using infrared data, yet still see similar excess,
this would be evidence for a larger effect from magnetic fields (or
another hidden parameter) than currently thought. Unfortunately, the
errors on our radii do not allow for a robust claim of this nature,
but it is an interesting avenue for the field.

The components of our new MEBs do not seem to converge towards the
standard 5 Gyr isochrone as they approach the fully-convective region.
In fact, our lowest mass star, which has a mass error bar that
straddles the fully-convective boundary, is the most inflated of the
six components we have measured. The lower panel of
Figure~\ref{fig:mr-rel} illustrates this inflation more clearly by
showing the radius anomaly $R_{\rm obs}/R_{\rm model}$ as a function
of mass, as computed with the standard $5$ Gyr isochrone. The errors
on the radius anomaly include the observed error on the radius and the
observed error on the mass (which propagates into the value of $R_{\rm
  model}$), added in quadrature.  The spread in radii at a given mass
is clearer here, and we discuss why stars of the same mass could be
inflated by different amounts in Section~\ref{sec:mrprel} by
considering their rotational velocities.

A comparison of the measured radii of all known MEBs to the model
isochrones shown in Figure~\ref{fig:mr-rel} might lead one to invoke
young ages for most of the systems, because stars with
$M_{\star}\lesssim0.7\msun$ are still contracting onto the pre-main
sequence at an age $\lesssim 200$ Myr and therefore have larger
radii. While young stars exist in the solar neighbourhood (as shown by
e.g. \citet{Jef93} who found an upper limit of 10-15 young stars
within 25pc), it is highly unlikely that all of the known MEBs are
young.  Indeed, the derived surface gravities for our MEBs are
consistent with older main-sequence stars. We see emission of
H$\alpha$ in all three systems, which can be an indicator of youth,
but close binary systems are known to exhibit significantly more
activity than wide binaries or single stars of the same spectral type
(see e.g. \citealt{Shk10}). We therefore do not have independent
evidence to strongly associate the inflated radii of our MEBs with
young ages.

\subsection{The mass-$\rm T_{\rm eff}$ diagram}\label{sec:mtrel}
\begin{figure*}
  \centering
  \includegraphics[width=\textwidth]{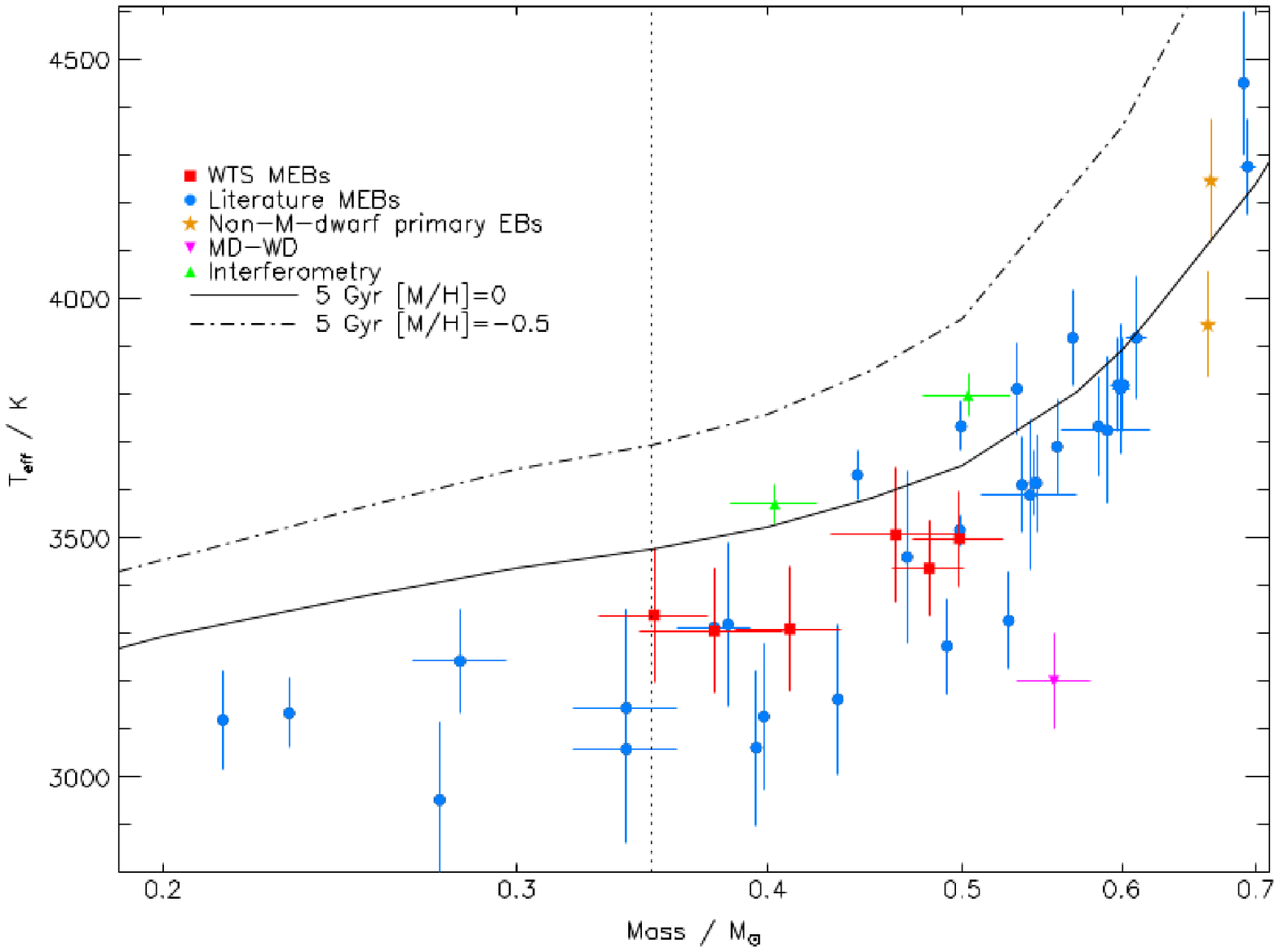}
  \caption{The mass-$\rm T_{\rm eff}$ diagram for low mass stars. Two
    different metallicity isochrones from the \citet{Bar98} 1 Gyr
    models are over-plotted to show the effect of decreasing
    metallicity. The vertical dotted line marks the fully-convective
    boundary \citep{Chab97}.  The data used in this plot are given in
    Table~\ref{tab:mrptdata}.}
  \label{fig:mt-rel}
\end{figure*}

As discussed in Section~\ref{sec:intro}, there is some evidence for a
radius-metallicity correlation \citep{Ber06,Lop07} amongst
M-dwarfs. Model values for effective temperatures depend on model
bolometric luminosities, which are a function of
metallicity. Metal-poor stars are less opaque so model luminosities
and effective temperatures increase while the model radii shrink by a
small amount \citep{Bar98}. Figure~\ref{fig:mt-rel} shows our MEBs in
the mass-$\rm T_{\rm eff}$ plane plus the same literature systems from
Figure~\ref{fig:mr-rel} where effective temperatures are
available. The two lines show the standard 5 Gyr isochrone of the
\citet{Bar98} models for solar metallicity stars (solid line) and for
metal-poor stars (dot-dash line).

The large errors in the mass-$\teff$ plane for M-dwarfs
mean that it is not
well-constrained. Section~\ref{sec:indices} has already highlighted
some of difficulties in constraining effective temperatures and
metallicities for M-dwarfs, but one should also note that effective
temperatures reported in the literature are determined using a variety
of different methods, e.g.  broad-band colour indices, spectral
indices, or model atmosphere fitting using several competing radiative
transfer codes. It also involves a number of different spectral type -
$\teff$ relations, and as \citet{Rey11} have demonstrated, these can
differ by up to $500$ K for a given M-dwarf subclass.

While the intrinsic scatter in the effective temperatures at a given
mass may be caused by metallicity effects, the overall trend is that
models predict temperatures that are too hot compared to observed
values, especially below $0.45\msun$. Our new MEBs, which we
determined to have metallicities consistent with the Sun, also conform
to this trend.  Furthermore, several studies of the inflated CM Dra
system have found it to be metal-poor \citep{Vit97,Vit02}, whereas
models would suggest it was metal-rich for its mass, based on its
cooler temperature and larger radius (see Table~\ref{tab:mrptdata} for
data). In this case, the very precisely measured inflated radius of CM
Dra cannot be explained by a high metallicity effect. In fact, the
tentative association of two of our new MEBs with the slightly
metal-poor young-old disk population defined by \citet{Leg92}, would
also make it difficult to explain their inflated radii using the
metallicity argument.

The scatter in the mass-$\teff$ plane can also arise from spot
coverage due to the fact that very spotty stars have cooler effective
temperatures at a given mass, and consequently larger radii for a
fixed luminosity. Large spot coverage fractions are associated with
high magnetic activity, which is induced by fast rotational
velocities. Table~\ref{tab:dimensions} gives the synchronous
rotational velocities of the stars in our MEBs along with their
theoretical timescales for tidal circularisation and
synchronisation. Among our new systems, 19c-3-01405 contains the
slowest rotating stars ($\sim 4$ km/s) on account of its longer
orbital period, and its components have stellar radii that are the
most consistent with the standard 5 Gyr model. The other faster
rotating stars in our MEBs have radii that deviate from the model by
more than $1\sigma$. We discuss this tentative trend between radius
inflation and rotational velocity (i.e. orbital period, assuming the
systems are tidally-locked) in the next section.
 
\subsection{A mass-radius-period relationship?}\label{sec:mrprel}
\begin{figure*}
  \centering
  \includegraphics[width=\textwidth]{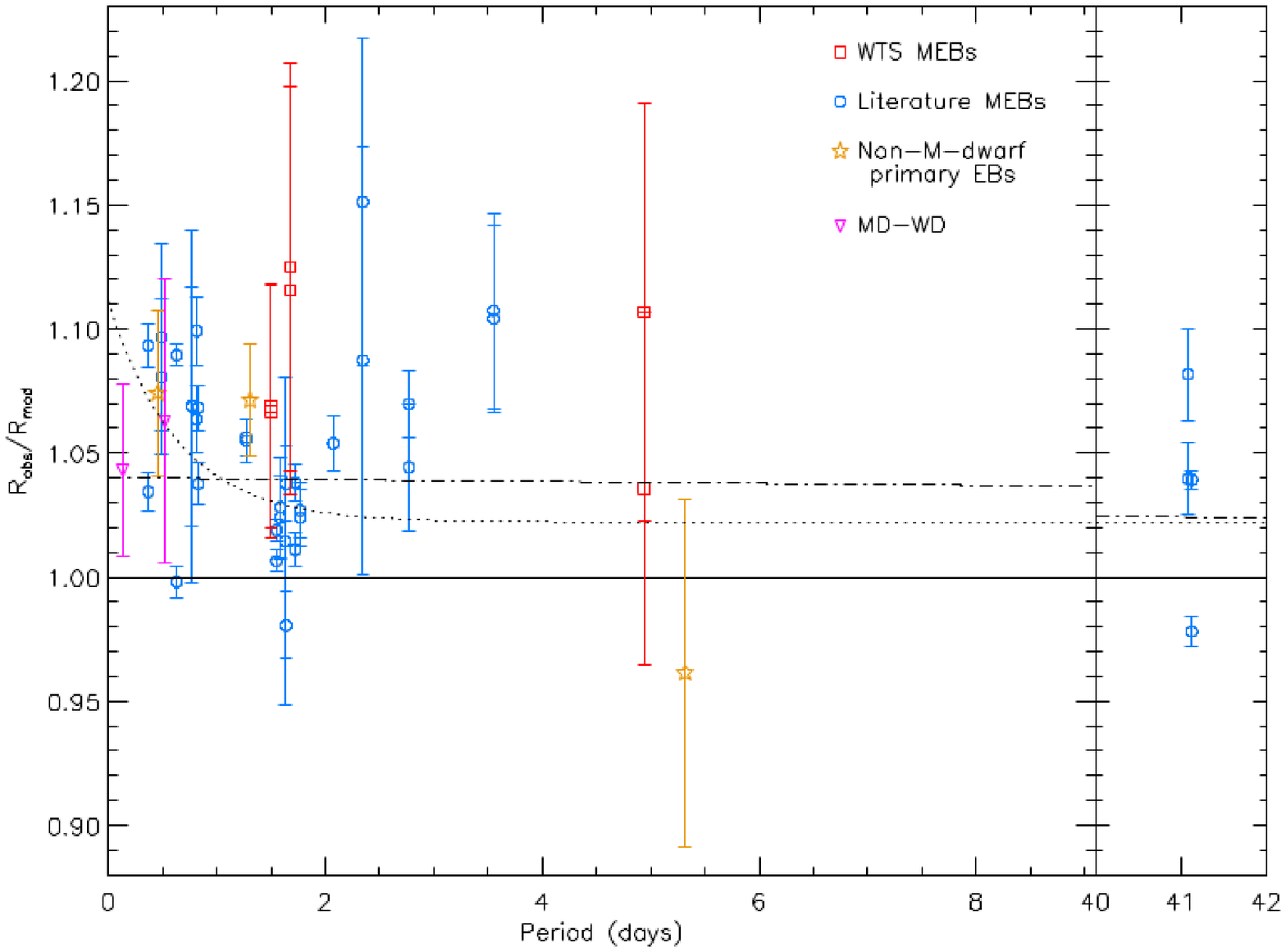}
  \caption[The Radius Anomaly as a Function of Orbital Period]{The
    radius anomaly as a function of orbital period using the 5 Gyr
    solar-metallicity isochrone from the \citet{Bar98} models. Our new
    MEBs are shown by the red open squares. Literature radius
    anomalies with radius errors $<6.5\%$ are also plotted. The errors
    are a quadrature sum of the measured radius error plus a
    propagated error from the observed mass which determines the model
    radius. The dashed and dotted lines show the best-fit from a
    straight-line and exponentially decaying model to the data,
    respectively. The coefficients and goodness of fit for these fits
    are given in Table~\ref{tab:linefit}.  The data used in this plot
    are given in Table~\ref{tab:mrptdata}.}
  \label{fig:pr-rel}
\end{figure*}

In a recent paper, \citet{Kra11a} presented six new MEBs with masses
between $0.38-0.59\msun$ and short orbital periods spanning $0.6-1.7$
days. Their measurements combined with existing literature revealed
that the mean radii of stars in systems with orbital periods less than
$1$ were different at the $2.6\sigma$ level to those at longer
periods. Those with orbital periods $<1$ day were systematically
larger than the predicted radii by $4.8\pm1\%$, whereas for periods
$>1.5$ days the deviation from the \citet{Bar98} models are much
smaller ($1.7\pm0.7\%$). The implication is that a very short orbital
period, i.e. very high level of magnetic activity, leads to greater
radius inflation, and one then expects the level of
  radius inflation to decrease at longer
  periods. Figure~\ref{fig:pr-rel} shows the radius anomaly
($R_{obs}/R_{model}$) as a function of period for our new MEBs plus
literature values whose reported errors are compatible with our own
measurements ($\sigma_{M_{\rm obs}}<6\%$ and $\sigma_{R_{\rm
    obs}}<6.5\%$). We used the 5 Gyr, solar metallicity isochrone from
the \citet{Bar98} models, with $L_{\rm mix}=H_{\rm P}$, to derive the
radius anomalies. The models were linearly interpolated onto a finer
grid with intervals of $0.0001\msun$, and the model photospheric radii
were calculated using $R_{\rm model}=\sqrt{L_{\rm model}/4\pi\sigma
  T_{\rm eff,model}^{4}}$.

Despite the small sample, we have performed an error weighted
statistical analysis of the period distribution, including our new
measurements, to compare to the unweighted analysis presented in
\citet{Kra11a}. Table~\ref{tab:stats} reports the weighted mean
($\bar{\mu}$) and weighted sample standard deviation ($\sigma$) of the
radius anomaly for three different period ranges: i) all periods, ii)
periods $\leq1$ day and, iii) periods $>1$ day. The boundary between
the `short' and `long' period samples was chosen
initially to match the analysis by \citet{Kra11a}. A
T-test using the weighted mean and variances of the short and long
period samples shows that their mean radii are distinct populations at
a $4.0\sigma$ significance, in support of \citeauthor{Kra11a}'s
findings. However, the significance level is strongly
  dependent on the chosen period boundary, and is skewed by the
  cluster of very precisely measured values near $1.5$ days. For
  example, a peak significance of $4.8\sigma$ is found when dividing
  the sample at $1.5$ days, but sharply drops to $\sim1\sigma$ for
  periods of $1.7$ days or longer. At short periods, it rises
  gradually towards the peak from $1\sigma$ at $0.3$ days.

\begin{table}
  \centering
  \begin{tabular}{cccc}
    \hline
    \hline
    Period&$\bar{\mu}$&$\pm \frac{\sigma}{\sqrt{N}}$&$\sigma$\\
    \hline
    All&103.6\%&0.5\%&3.2\%\\
    \hline
    $P\leq1.0$&106.1\%&0.9\%&3.5\%\\
    \hline
    $P>1.0$&102.6\%&0.4\%&2.4\%\\
    \hline
  \end{tabular}
  \caption{A statistical analysis of the mean radius inflation for
    different period ranges. $\sigma$ is the weighted sample standard deviation.}
  \label{tab:stats}
\end{table}

Instead, we have attempted to find a very basic
mathematical description for any correlation between
radius inflation and orbital period, but we appreciate our efforts are
hampered by small number statistics. We fitted the distribution of the
radius anomaly as a function of period, using first a
linear model and then as an exponentially decaying
function. We used the {\sc idl} routine {\sc mpfitfun} to determine an
error weighted best-fit and the $1\sigma$ errors of the model
parameters. The results are reported in Table~\ref{tab:linefit} and
the best-fit models are over-plotted in
Figure~\ref{fig:pr-rel}, but neither model is a good fit (although the
exponential fairs moderately better). While there is
  marginal evidence for greater inflation in the shortest period
  systems, we find that the expected convergence towards theoretical
  radius values for longer period, less active systems is not
  significantly supported by the available observation data.

\begin{table*}
  \centering
  \begin{tabular}{lcccccc}
    \hline
    \hline
    Model&$a_{0}$&$a_{1}$&$a_{2}$&$\chi^{2}$&DOF&$\chi_{\nu}^{2}$\\
    \hline
    $R_{\rm obs}/R_{\rm mod}=a_{0}+a_{1}P$&$1.0401\pm0.0017$&$-0.000386\pm0.000086$&--&490.5&46&10.7\\
    $R_{\rm obs}/R_{\rm mod}=a_{0}+a_{1}e^{a_{2}P}$&$1.0221\pm0.0027$&$0.089\pm0.015$&$-1.57\pm0.33$&401.6&45&8.9\\
    \hline
  \end{tabular}
  \caption{Results from an error weighted modelling of the radius anomaly as a function of
    period. $a_{i}$ are the coefficients of the models and $P$ is the
    orbital period in days. Neither of these simple models provide a
    statistically good fit, indicating a more complex relationship
    between the radius anomaly and orbital period.}
  \label{tab:linefit}
\end{table*}

There are two pertinent observations worth addressing, namely the
low-mass eclipsing binaries LSPM J1112+7626 and Kepler-16
\citep{Irw11,Doy11,Ben12}, which were announced after the
\citet{Kra11a} study. These systems significantly extended the
observed orbital period range, with almost identical 41-day orbital
periods, and both containing one fully-convective component
($M_{\star}\sim0.35\msun$, \citealt{Chab97}) and one partially
convective component (see Table~\ref{tab:mrptdata}). The radius
inflation differs significantly between these two systems, as shown on
the right-hand side of Figure~\ref{fig:pr-rel}. While the more
massive, partially-convective component of Kepler-16 is well-described
by the 1 Gyr model isochrone \citet{Bar98} (see
Figure~\ref{fig:mr-rel}), the other three stars suffer significant
radius inflation, with no obvious correlation between the amount of
inflation and the masses, even though one of them is a
partially-convective star. This residual inflation, particularly for
the fully-convective stars at long periods, may pose a challenge to
the magnetic activity hypothesis as the sole reason for discrepancies
between models and observations, especially given the extremely
high-quality measurements of Kepler-16. However, one should note that
other studies have suggested that the presence of a strong magnetic
field can alter the interior structure of a low-mass star, such that
is pushes the fully-convective mass limit for very active stars to
lower values \citep{Mul01,Chab07}, so these stars may still suffer
from a significant inhibition of convective flow.

The radius anomaly raises concern over the usefulness of the known
MEBs in calibrating models for the evolution of singular M-dwarf stars
that are the favoured targets of planet-hunting surveys searching for
habitable worlds. \citet{Kra11a} argue that the high-activity levels
in very close MEBs make them poor representatives of typical single
low-mass stars and that the observed radius discrepancies should not
be taken as an indictment of stellar evolution models. However, we
have seen that radius inflation remains in MEBs systems with low
magnetic activity and furthermore, the inflated
  components of LSPM J1112+7626 do not exhibit H$\alpha$ emission
that is typically associated with the high activity levels in MEBs
with inflated radii. \citet{Wes11} used H$\alpha$ emission as an
activity indicator to determine that the fraction of single, active,
early M-dwarfs is small ($<5\%$), but increases to $40-80\%$ for M4-M9
dwarfs. Yet, it may be that the amount of activity needed to inflate
radii to the measured values in MEBs is small and therefore below the
level where observable signatures appear in H$\alpha$ emission. This
would then question the reliability of H$\alpha$ emission as an
activity indicator, meaning the fraction of `active', single M-dwarfs
may be even higher than the \citet{Wes11} study. Given that these very
small stars are a ripe hunting-ground for Earth-size planets, we must
be able to constrain stellar evolution models in the presence of
magnetic activity if we are to correctly characterise planetary
companions. We note that even the very precisely-calibrated
higher-mass stellar evolution models \citep{And91,Tor10} do not
reproduce the radii of active stars accurately (see \citet{Mor09} who
found $4-8\%$ inflation in a G7+K7 binary with a 1.3 day orbit).

In order to establish a stringent constraint on the relationship
between mass, period and radius, we need further measurements of
systems that i) include `active' and `non-active' stars that span the
fully-convective and partially-convective mass regimes, and ii) a
better sampled range of orbital periods beyond 5 days to explore
systems that are not synchronised. We may ultimately find that
activity does not account for the full extent of the radius anomaly,
and as suggested by \citet{Irw11}, perhaps the equation of state for
low-mass stars can still be improved. On the other hand, perhaps the
importance of tidal effects between M-dwarfs in binaries with wider
separations has been underestimated, as it has been shown that the
orbital evolution of M-dwarf binary systems is not well-described by
current models \citep{Nef12}.

\section{Conclusions}
In this paper, we have presented a catalogue of $16$ new low-mass,
detached eclipsing binaries that were discovered in the WFCAM Transit
Survey. This is the first time dynamical measurements of M-dwarf EBs
have been detected and measured primarily with infrared data. The
survey light curves are of high quality, with a per epoch photometric
precision of $3-5$ mmag for the brightest targets ($J\sim13$
mag), and a median RMS of
  $\lesssim1\%$ for $J\lesssim16$ mag. We have reported the
characterisation of three of these new systems using follow-up
spectroscopy from ground-based $2-4$ m class telescopes. The three
systems ($i=16.7-17.6$ mag) have orbital periods in the range
$1.5-4.9$ days, and span masses $0.35-0.50\msun$ and radii
$0.38-0.50\rsun$, with uncertainties of $\sim 3.5-6.4\%$ in mass and
$\sim 2.7-5.5\%$ in radius. Two of the systems may be associated with
the young-old disk population as defined by \citet{Leg92} but our
metallicity estimates from low-resolution spectra do not confirm a
non-solar metallicity.

The radii of some of the stars in these new systems are significantly
inflated above model predictions ($\sim3-12\%$). We analysed their
radius anomalies along with literature data as a function of the
orbital period (a proxy for activity). Our
  error-weighted statistical analysis revealed marginal evidence for
  greater radius inflation in very short orbital periods $<1$ day, but
  neither a linear nor exponentially decay model produced a
  significant fit to the data. As a result, we found no statistically
  significant evidence for a correlation between the radius anomaly
  and orbital period, but we are limited by the small sample of
  precise mass and radius measurements for low-mass stars. However, it
  is clear that radius inflation exists even at longer orbital periods
  in systems with low (or undetectable) levels of magnetic activity. A
  robust calibration of the effect of magnetic fields on the radii of
  M-dwarfs is therefore a key component in our understanding of these
  stars. Furthermore, it is a limiting factor in characterising the
  planetary companions of M-dwarfs, which are arguably our best
  targets in the search for habitable worlds and the study of other
  Earth-like atmospheres.

More measurements of the masses, radii and orbital periods of M-dwarf
eclipsing binaries, spanning both the fully convective regime and
partially convective mass regime, for active and non-active stars,
across a range of periods extending beyond 5 days, are necessary to
provide stringent observational constraints on the role of activity in
the evolution of single low-mass stars. However, the influence of
spots on the accuracy to which we can determine the radii from light
curves will continue to impede these efforts, even in the most careful
of cases (see e.g. \citealt{Mor10,Irw11}).

This work has studied only one third of the M-dwarfs in the WFCAM
Transit Survey. Observations are on-going and we expect our catalogue
of M-dwarf eclipsing binaries to increase. This forms part of the
legacy of the WTS and will provide the low-mass star community with
high-quality MEB light curves. Furthermore, the longer the WTS runs,
the more sensitive we become to valuable long-period, low-mass
eclipsing binaries. These contributions plus other M-dwarf surveys,
such as MEarth and PTF/M-dwarfs, will ultimately provide the
observational calibration needed to anchor the theory of low-mass
stellar evolution.

\section*{Acknowledgements}
We would like to thank I. Baraffe for providing the model magnitudes
for our SED fitting in the appropriate filters, and S. Aigrain for the
use of the {\sc occfit} transit detection algorithm. This work was
greatly improved by several discussions with J. Irwin, as well as J.
Southworth, K. Stassun and R. Jeffries. We thank the referee for their
insightful comments which have improved this work and we also thank C.
del Burgo for his comments on the original manuscript. JLB acknowledges
the support of an STFC PhD studentship during part of this research.
We thank the members of the WTS consortium and acknowledge the support
of the RoPACS network. GK, BS, PC, NG, and HS are supported by RoPACS,
while JLB, BN, SH, IS, DP, DB, RS, EM and YP have received support
from RoPACS during this research, a Marie Curie Initial Training
Network funded by the European Commission’s Seventh Framework
Programme. NL is supported by the national program AYA2010-19136
funded by the Spanish ministry of science and innovation. Finally, we
extend our thanks to the fantastic team of TOs and support staff at
UKIRT, and all those observers who clicked on U/CMP/2.

The United Kingdom Infrared Telescope is operated by the Joint
Astronomy Centre on behalf of the Science and Technology Facilities
Council of the U.K. This article is based on observations made with
the INT, WHT operated on the island of La Palma by the ING in the
Spanish Observatorio del Roque de los Muchachos, and with the IAC80 on
the island of Tenerife operated by the IAC in the Spanish Observatorio
del Teide. This research uses products from SDSS DR7. Funding for the
SDSS and SDSS-II has been provided by the Alfred P. Sloan Foundation,
the Participating Institutions, the National Science Foundation, the
U.S. Department of Energy, the National Aeronautics and Space
Administration, the Japanese Monbukagakusho, the Max Planck Society,
and the Higher Education Funding Council for England. The SDSS Web
Site is http://www.sdss.org/. This publication makes use of data
products from the Two Micron All Sky Survey, which is a joint project
of the University of Massachusetts and the Infrared Processing and
Analysis Center/California Institute of Technology, funded by the
National Aeronautics and Space Administration and the National Science
Foundation. This work also makes use of NASA’s Astrophysics Data
System (ADS) bibliographic services, and the SIMBAD database, operated
at CDS, Strasbourg, France. {\sc iraf} is distributed by the National
Optical Astronomy Observatory, which is operated by the Association of
Universities for Research in Astronomy (AURA) under cooperative
agreement with the National Science Foundation.

\bibliographystyle{mn2e}
\bibliography{referencesjlb}{}

\clearpage
\appendix
\section{The full WTS 19hr field M-dwarf eclipsing binary sample}
In Table~\ref{tab:others}, we present the periods, epochs, effective
temperatures, $J$-band and $i$-band magnitudes of the $13$ remaining
19hr detached, well-sampled M-dwarf eclipsing binaries found with this
study ($J\leq16$ mag). The temperatures are based on
the SED fitting described in Section~\ref{sec:SED} and may be
under-estimated. The periods and epochs are based only on least-square
fitting which under-estimates the errors. These results are accurate
to $\sim30$ minutes and we recommend to anyone planning to observe
these objects in a time critical manner that they check these values
themselves with the light curve data provided with this paper. Note
that 19g-4-02069 is the subject of a near future publication (Nefs et
al. \emph{in prep.}) using RVs follow-up already obtained with
GNIRS/GEMINI. The phase-folded light curves are shown in
Figures~\ref{fig:others}.  and~\ref{fig:others2}, and
  the light curve data are provide in Table~\ref{tab:wts_lcs_other}.
\begin{table*}
\centering
\begin{tabular}{lrrrrrrrrr}
  \hline
  \hline
  Name&RA&Dec&$N_{\rm epochs}$&RMS&$P$&$T_{0}$&$J$ (Vega)&$i$ (Vega)&$\rm T_{\rm eff,SED}$\\
  &(deg)&(deg)&&(mag)&(days)&(HJD)&(mag)&(mag)&(K)\\
  \hline
  19a-1-02980&292.71276&36.312725&$893$&5.8&$2.103525$&$2454318.65422$&$14.861\pm0.004$&$16.166\pm0.004$&$3946\pm100$\\
  19c-3-08647&294.30659&36.815037&$893$&15.0&$0.867466$&$2454318.50614$&$14.812\pm0.004$&$16.171\pm0.004$&$3883\pm100$\\
  19c-4-11480&293.81149&36.902880&$893$&20.4&$0.681810$&$2454317.89071$&$15.850\pm0.006$&$17.208\pm0.007$&$3946\pm100$\\
  19d-2-07671& 294.58622&36.386467&$891$&48.9&$0.614540$&$2454317.99692$&$15.971\pm0.007$&$17.101\pm0.007$&$4209\pm100$\\
  19d-2-09173&294.50246&36.365239&$891$&22.4&$3.345469$&$2454320.15668$&$15.185\pm0.005$&$16.343\pm0.005$&$4209\pm100$\\
  19e-2-02883&293.32813&36.241312&$898$&10.6&$0.810219$&$2454317.90290$&$15.976\pm0.007$&$17.272\pm0.007$&$3946\pm100$\\
  19f-1-07389&292.89403&36.143865&$904$&18.3&$0.269868$&$2454317.97411$&$15.504\pm0.005$&$16.575\pm0.005$&$4209\pm100$\\
  19f-4-05194&292.81253&36.590539&$904$&35.0&$0.589530$&$2454318.10730$&$16.013\pm0.007$&$17.070\pm0.006$&$4209\pm100$\\
  19g-1-13215&293.63655&36.249009&$898$&10.2&$2.843515$&$2454318.34495$&$15.985\pm0.007$&$17.589\pm0.008$&$3374\pm100$\\
  19g-2-08064&294.16931&36.162723&$898$&14.8&$1.720410$&$2454317.94781$&$14.466\pm0.003$&$15.596\pm0.004$&$4209\pm100$\\
  19g-4-02069&293.76480&36.521247&$898$&11.2&$2.441759$&$2454321.78532$&$14.843\pm0.004$&$16.911\pm0.006$&$3054\pm100$\\
  19h-2-00357&294.66466&36.272874&$885$&8.3&$7.004082$&$2454320.79766$&$15.531\pm0.005$&$16.808\pm0.006$&$3946\pm100$\\
  19h-2-01090&294.62103&36.262345&$886$&11.5&$5.285051$&$2454322.78131$&$15.681\pm0.006$&$16.843\pm0.006$&$4209\pm100$\\
  \hline
\end{tabular}
\caption{The first release of the WTS M-dwarf Eclipsing Binary
  Catalogue detailing the remaining MEBs in the WTS 19hr field with
  $J\leq16$ mag that are not characterised in this paper. Note
  that 19g-4-02069 is the subject for a near future publication by Nefs et
  al. (\emph{in prep.}) using RV follow-up from GNIRS/GEMINI. Please see
  appendix text for caveats on the quoted ephemerides.}
\label{tab:others}
\end{table*}

\begin{table}
  \centering
  \begin{tabular}{@{\extracolsep{\fill}}lccc}
    \hline
    \hline
    Name&HJD&$J_{\rm WTS}$&$\sigma_{J_{\rm WTS}}$\\
    &&(mag)&(mag)\\
    \hline
    19a-1-02980&2454317.82863842&14.846761&0.004826\\
    19a-1-02980&2454317.84010834&14.844511&0.004894\\
    ...&...&...&...\\
    \hline
  \end{tabular}
  \caption{The WTS $J$-band light curves for the remainder of the WTS
    MEB catalogue given in Table~\ref{tab:others} Magnitudes are given in the WFCAM system.
    \citet{Hod09} provide conversions for other systems. The errors,
    $\sigma_{J}$, are estimated using a standard noise model,
    including contributions from Poisson noise in the stellar counts,
    sky noise, readout noise and errors in the sky background
    estimation.
    (This table is published in full in the online journal
    and is shown partially here for guidance regarding its form and content.)}
  \label{tab:wts_lcs_other}
\end{table}

\begin{figure*}
\centering
\includegraphics[width=\textwidth]{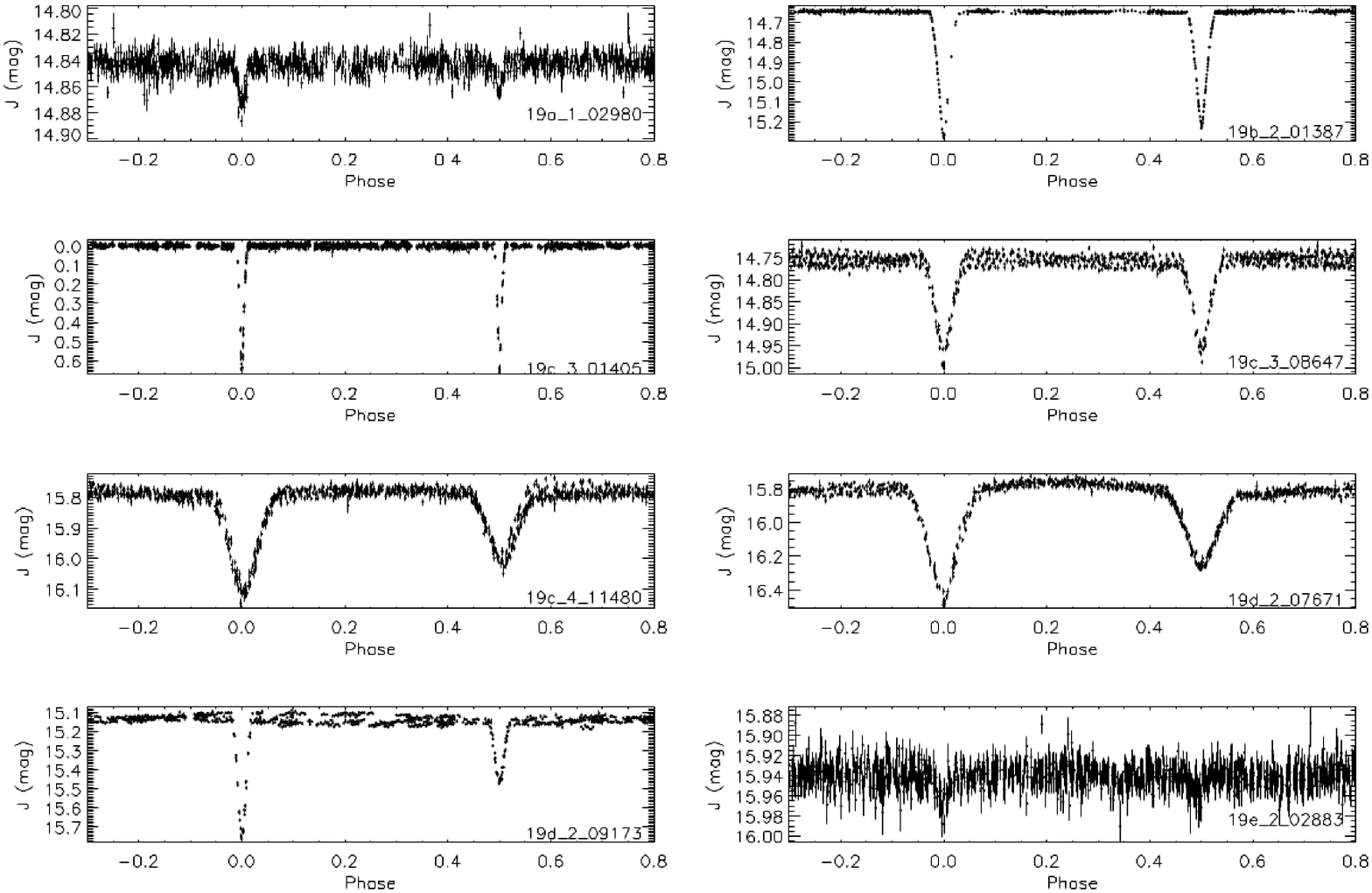}
\caption{Phase-folded light curves of the MEBs discovered in the WTS
  19hr field with $J\leq16$ mag...}
\label{fig:others}
\end{figure*}

\begin{figure*}
\centering
\includegraphics[width=\textwidth]{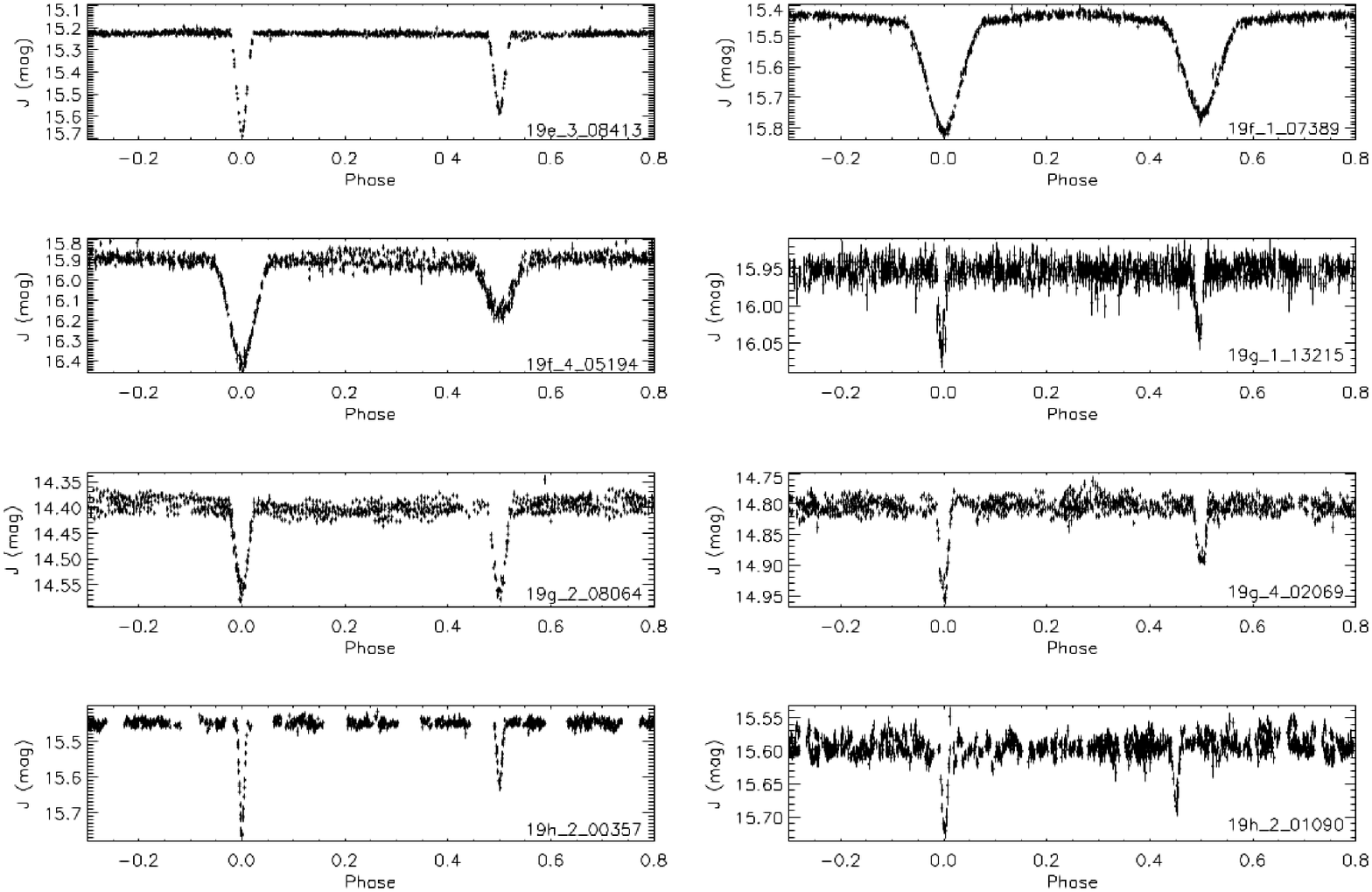}
\caption{cont... Phase-folded light curves of the MEBs discovered in
  the WTS 19hr field with $J\leq16$ mag.}
\label{fig:others2}
\end{figure*}

Table~\ref{tab:mrptdata} contains the literature data used to create
Figures~\ref{fig:mr-rel},~\ref{fig:mt-rel} and~\ref{fig:pr-rel}. The
literature data was selected with the following filters: mass errors
$<6.4\%$ and radius errors $<5.5\%$ (comparable to or better than the
errors we presented for the three characterised MEBs in this paper),
and in the range $0.19\leq M_{\star}\leq0.71$ and $0.19\leq
R_{\star}\leq0.71$.

\begin{table*}
  \centering
   \begin{tabular}{lrrrrrrrr}
    \hline
    \hline
    Name&Period&Mass&$\sigma_{M}$&Radius&$\sigma_{R}$&$\rm T_{\rm eff}$&$\sigma_{\rm T_{\rm
        eff}}$&Ref\\
    &(days)&$(\msun)$&$(\msun)$&$(\rsun)$&$(\rsun)$&(K)&(K)&\\
    \hline
    MEBs&&&&&&&&\\
    NSVS01031772A&0.368&0.5428&0.0028&0.5260&0.0028&3614.1&67.2&(1)\\
    NSVS01031772B&0.368&0.4982&0.0025&0.5087&0.0031&3515.6&32.5&(1)\\
    GUBooA&0.489&0.6100&0.0071&0.6230&0.0163&3917.4&128.3&(1)\\
    GUBooB&0.489&0.5990&0.0061&0.6200&0.0203&3810.7&133.9&(1)\\
    MG1-1819499A&0.6303135&0.557&0.001&0.569&0.002&3690.0&100.0&(2)\\
    MG1-1819499B&0.6303135&0.535&0.001&0.500&0.003&3610.0&100.0&(2)\\
    GJ3236A&0.77126&0.376&0.016&0.3795&0.0084&3312.0&110.0&(3)\\
    GJ3236B&0.77126&0.281&0.015&0.300&0.015&3242.0&108.0&(3)\\
    YYGemA&0.814&0.5974&0.0047&0.6196&0.0057&3819.4&98.0&(1)\\
    YYGemB&0.814&0.6009&0.0047&0.6035&0.0057&3819.4&98.0&(1)\\
    MG1-116309A&0.8271425&0.567&0.002&0.552&0.004&3917.4&100.5&(2)\\
    MG1-116309B&0.8271425&0.532&0.002&0.532&0.004&3810.7&97.8&(2)\\
    CMDraA&1.268&0.2310&0.0009&0.2534&0.0019&3133.3&73.0&(1)\\
    CMDraB&1.268&0.2141&0.0009&0.2396&0.0015&3118.9&102.2&(1)\\
    MG1-506664A&1.5484492&0.584&0.002&0.560&0.001&3732.5&104.6&(2)\\
    MG1-506664B&1.5484492&0.544&0.002&0.513&0.001&3614.1&101.3&(2)\\
    MG1-78457A&1.5862046&0.5270&0.0019&0.505&0.008&3326.6&101.1&(2)\\
    MG1-78457B&1.5862046&0.491&0.002&0.471&0.009&3273.4&99.5&(2)\\
    LP133-373A&1.6279866&0.34&0.02&0.330&0.014&3144.0&206.0&(4)\\
    LP133-373B&1.6279866&0.34&0.02&0.330&0.014&3058.0&195.0&(4)\\
    MG1-646680A&1.6375302&0.499&0.002&0.457&0.006&3732.5&51.9&(2)\\
    MG1-646680B&1.6375302&0.443&0.002&0.427&0.006&3630.8&50.5&(2)\\
    MG1-2056316A&1.7228208&0.4690&0.0021&0.441&0.002&3459.4&179.8&(2)\\
    MG1-2056316B&1.7228208&0.382&0.002&0.374&0.002&3318.9&172.5&(2)\\
    KOI126B&1.76713&0.2413&0.0030&0.2543&0.0014&--&--&(5)\\
    KOI126C&1.76713&0.2127&0.0026&0.2318&0.0013&--&--&(5)\\
    HIP96515Aa&2.3456&0.59&0.03&0.64&0.01&3724.0&154.0&(6)\\
    HIP96515Ab&2.3456&0.54&0.03&0.55&0.03&3589.0&157.0&(6)\\
    CUCncA&2.771&0.4333&0.0017&0.4317&0.0052&3162.3&156.7&(1)\\
    CUCncB&2.771&0.3980&0.0014&0.3908&0.0095&3126.1&154.9&(1)\\
    1RXSJ154727A&3.5500184&0.2576&0.0085&0.2895&0.0068&--&--&(7)\\
    1RXSJ154727B&3.5500184&0.2585&0.0080&0.2895&0.0068&--&--&(7)\\
    LSPMJ1112A&41.03236&0.3946&0.0023&0.3860&0.005&3061.0&162.0&(8)\\
    LSPMJ1112B&41.03236&0.2745&0.0012&0.2978&0.005&2952.0&163.0&(8)\\
    Kepler16A&41.079220&0.6897&0.0035&0.6489&0.0013&4450&150&(9)\\
    Kepler16B&41.079220&0.20255&0.00066&0.22623&0.00059&--&--&(9)\\
    \hline
    Non-M-dwarf primary EBs&&&&&&&&\\
    NGC-2204-S892B&0.4520000&0.6621&0.0050&0.6800&0.0203&3944.6&110.5&(1)\\
    IM-VirB&1.3090000&0.6644&0.0048&0.6809&0.0131&4246.2&129.0&(1)\\
    RXJ0239B&2.0720160&0.693&0.006&0.703&0.002&4275.0&109.0&(10)\\
    \hline
    MD-WD EBs&&&&&&&&\\
    RXJ2130&0.5210356&0.555&0.023&0.553&0.017&3200.0&100.0&(10)\\
   \hline
    Interferometry&&&&&&&&\\
    GJ411  &--&0.403&0.020&0.393&0.008&3570.0&42.0&(11)\\
    GJ380  &--&0.670&0.033&0.605&0.020&--&--&(11)\\
    GJ887  &--&0.503&0.025&0.459&0.011&3797.0&45.0&(12)\\
\hline
\end{tabular}
\caption{Literature values for systems used in
  Figures~\ref{fig:mr-rel},~\ref{fig:mt-rel} and~\ref{fig:pr-rel}
  with mass errors $<6.4\%$ and radius errors $<5.5\%$, in the range
  $0.19\leq M_{\star}\leq0.71$ and $0.19\leq
  R_{\star}\leq0.71$. Temperatures are given when available in the
  literature, but those without are not included in
  Figure~\ref{fig:mr-rel}. There are no rotation periods given for the
  interferometric measurements therefore these are excluded from Figure~\ref{fig:pr-rel}.
  References:
  (1) DEBCat and references therein (www.astro.keele.ac.uk/jkt/debcat/), 
  (2) \citet{Kra11a}, 
  (3) \citet{Irw09b}, 
  (4) \citet{Vac07}, 
  (5) \citet{Cart11}, 
  (6) \citet{Hue09}, 
  (7) \citet{Har11}, 
  (8) \citet{Irw11},
  (9) \citet{Doy11},
  (10) \citet{Kni11} and references therein, 
  (11) \citet{seg03}, 
  (12) \citet{Dem09}.}
\label{tab:mrptdata}
\end{table*}

\label{lastpage}
\end{document}